\documentclass[fleqn, usenatbib]{mnras}

\usepackage{newtxtext,newtxmath}

\usepackage[T1]{fontenc}

\usepackage{graphicx}	
\usepackage{amsmath}	
\usepackage{mathrsfs}
\usepackage[flushleft]{threeparttable}
\usepackage{siunitx}
\usepackage{booktabs, caption}
\usepackage{float}
\usepackage{pdflscape}
\usepackage{ulem}
\usepackage{physics}

\newcommand{\GG}[1]{}

\title[PAH Kinematics in local Seyferts]{GATOS XIV: The first direct kinematic evidence of dusty outflows from AGN via PAH kinematics of local Seyfert galaxies with JWST}

\author[F. R. Donnan et al.]{
Fergus R. Donnan,$^{1,2}$\thanks{E-mail: fdonnan@ucsd.edu}
Ismael Garc{\'i}a-Bernete,$^{3}$
Dimitra Rigopoulou,$^{1,4}$
Almudena Alonso-Herrero,$^{3}$
\newauthor
Anelise Audibert,$^{5,6}$
Enrica Bellocchi,$^{7,8}$
Andrew Bunker,$^{1}$
Steph Campbell,$^{9}$
Fran\c{c}oise Combes,$^{10}$
\newauthor
Richard Davies,$^{11}$
Tanio D\'{i}az-Santos,$^{12,4}$
Juan A. Fern\'andez-Ontiveros,$^{13}$
Poshak Gandhi,$^{14}$
\newauthor
Santiago Garc{\'i}a-Burillo,$^{15}$
Omaira Gonz\'alez-Mart\'in,$^{16}$
Erin K. S. Hicks,$^{17}$
Laura Hermosa Muñoz,$^{3}$
\newauthor
Sebastian F. Hoenig,$^{14}$
Masatoshi Imanishi,$^{18,19}$
Alvaro Labiano,$^{20}$
Nancy A. Levenson,$^{21}$
\newauthor
Miguel Pereira-Santaella,$^{22}$
Cristina Ramos Almeida,$^{5,6}$
Claudio Ricci,$^{23,24}$
Rogemar A. Riffel,$^{25}$
\newauthor
Daniel Rouan,$^{26}$
David Rosario,$^{9}$
Karin Sandstrom,$^{2}$
Taro T. Shimizu,$^{11}$
Marko Stalevski,$^{27,28}$
Niranjan Thatte,$^{1}$
\newauthor
Oscar Veenema,$^{1}$
Lulu Zhang,$^{29}$
\\
\\
(Affiliations can be found after the references)
}

\date{Accepted XXX. Received YYY; in original form ZZZ}

\pubyear{2024}

\begin{document}
\label{firstpage}
\pagerange{\pageref{firstpage}--\pageref{lastpage}}
\maketitle

\begin{abstract}

We present the first spatially resolved kinematic evidence for dust in the outflows of Active Galactic Nuclei (AGN). We utilise observations from JWST with NIRSpec IFU and MIRI MRS data of 10 local Seyferts and use Principal Component Analysis (PCA) tomography to extract the kinematics of Polycyclic Aromatic Hydrocarbon (PAH) features. PAHs comprise the smallest carbonaceous dust molecules in the Interstellar Medium (ISM), and produce emission features in the infrared providing the potential to measure kinematics. This is however challenging due to their broad shapes and variations in their intrinsic profile, prompting the need for techniques such as PCA tomography.
We find that the velocity of the PAHs is similar to the molecular gas as traced by the rotational transitions of H$_2$, where for NGC 5728 and NGC 7582, both disk and outflow are present. We detect the outflow in the kinematics of large and neutral PAHs, namely the 11.3 $\mu$m and 17 $\mu$m PAH features, where after subtracting the disk, the velocity field matches that of high-ionisation potential lines such as [NeVI] (7.65 $\mu$m, IP = 158 eV). Finally, we fail to detect kinematics of the 6.2 $\mu$m PAH due to an altered intrinsic profile while the the 3.3 $\mu$m PAH kinematics purely trace the circumnuclear disk. This suggests the PAHs in the outflow are more neutral and larger than in star-forming regions, consistent with PAH band ratios in previous studies of AGN.

\end{abstract}




\section{Introduction}

Active Galactic Nuclei (AGN), powered by accretion onto a supermassive black hole (SMBH), are known to play a key role in the evolution of galaxies \citep[e.g.][]{Kormendy2013, Heckman2014}, where vast quantities of energy are radiatively and kinematically ejected into the host galaxy \citep[e.g.,][]{Silk1998, Fabian2012, Combes2017}. 

The idea that AGN are dynamic and evolving systems that interact with the host galaxy, suggests that the early model of a static torus \citep{Antonucci1985, Antonucci1993}, to explain the presence or lack of presence of broad emission lines on AGN spectra, does not tell the full picture. More contemporary models \citep[See][for recent reviews.]{ RamosAlmeida2017, Honig2019} feature clumpy or two-phase media, with polar and toroidal dust \citep[e.g.][]{Nenkova2008a, Nenkova2008b, Honig2010b,  Honig2017, Stalevski2016} or that the torus itself is the result of a failed wind \citep[][]{Wada2012}.

The existence of dust in the polar direction, perpendicular to the torus, has long been postulated from ground based imaging \citep[][]{Braatz1993, Bock2000, Radomski2002, Radomski2008, Packham2005, Reunanen2010, Asmus2014, Asmus2016, Asmus2019, Stalevski2017} but particularly with the advent of interferometry in the infrared revealing polar structure \citep[][]{Raban2009,Honig2012, Honig2013, Lopez-Gonzaga2014, Tristram2014, Lopez-Gonzaga2016,  Leftley2018, Isbell2022, GamezRosas2022}. Interferometry probes dust on very small scales, comparable to the torus itself \citep[$\sim 10$s pc][]{Garcia-Burillo2019, Garcia-Burillo2021, Combes2019, Alonso-Herrero2021} and are thought to be dust driven from the inner regions of the torus by an AGN wind, alongside the molecular and ionised phases of the outflow. Larger scale dusty structures have also been detected with JWST/MIRI imaging after carefully removing contamination from emission lines \citep[][]{Haidar2024, Campbell2025, Haidar2026}. It is unclear however if this dust is simply heated by the AGN or actually launched by the AGN. Indeed \citet{Haidar2026} suggests AGN radiation alone is not sufficient to heat the dust and that shocks may play a role. Moreover, AGN outflows can entrain material from the host galaxy \citep[e.g.][]{Garcia-Bernete2021}, which may be necessary to launch dusty outflows. A key missing diagnostic of this dust, unlike the gas phase, are its kinematics.

Polycyclic Aromatic Hydrocarbons (PAHs) comprise the smallest carbonaceous dust molecules and consist of rings of carbon bonded with hydrogen. These molecules are radiatively excited by UV photons from young stars, where the subsequent de-excitation via the stretching and bending of C-C and C-H bonds results in a suite of broad emission features in the infrared (3-17 $\mu$m) \citep[e.g.][]{Tielens2008, Li2020}. While graphite and silicate dust grains produce a continuum that is unsuitable to measuring kinematics, the emission profiles of PAH molecules provide the possibility of measuring kinematics, but their broad shapes provide significant challenges \citep[][]{Donnan24b}.

Unlike in star-forming regions, PAHs exhibit low equivalent widths (EW) in AGN due to the strong infrared continuum in such sources but also destruction of PAH molecules from the AGN radiation field \citep[e.g.][]{Rapacioli2006, Schweitzer2006, Alonso-Herrero2014} or shocks \citep[e.g.][]{Micelotta2010}. Recent studies, taking advantage of the increased sensitivity and spatial resolution of JWST, have found that PAHs survive in the vicinity of AGN and in their outflows \citep[][]{ Garcia-Bernete2022b, Garcia-Bernete2022c, Lai2023, Garcia-Bernete2024b, Rigopoulou2024, Zhang2024, Donnelly2024, RamosAlmeida2025}. However, their properties appear to be altered, where neutral PAHs appear to survive while AGN have less ionised PAHs potentially suggesting preferential destruction of ionised PAHs. It is unclear however whether these PAHs are travelling within the AGN outflow itself or are simply residing in star-forming regions that are impacted by the outflow where the PAH properties become altered. To answer this question requires a kinematic analysis of the PAH features.

Measuring the kinematics of PAH features is challenging for two main reasons. Firstly, they are significantly broader than any Doppler shift in wavelength due to the velocity of the PAH molecules. Secondly, the intrinsic shape of the feature is not well defined, often appearing asymmetric, containing broad wings akin to a Lorentzian or Drude profile rather than a Gaussian Moreover the profile can change depending on the properties of the PAH population such as charge, size and the hardness of the radiation field \citep[e.g.][]{Peeters2002, Candian2015, Shannon2019}. This makes it difficult to establish the rest-frame emission to provide a reference to measure any deviations due to the velocity field of the PAHs.  

To overcome the above issues, \citet{Donnan24b} presented the first use of the technique of Principal Component Analysis (PCA) tomography to detect the signature of kinematics of the 3.3 $\mu$m PAH in three local LIRGs. The technique of PCA tomography was first presented in \citet{Steiner2009}. PCA tomography, when applied to a data cube containing a PAH feature, can detect the spectral signature of a blue/red shifted feature, using the information held within the spatial axes. This allows one to construct a velocity map of these features without modelling the spectra of each spaxel.

In this work we apply PCA tomography to a sample of nearby Seyfert galaxies, observed with JWST/MIRI MRS and JWST/NIRSpec IFU, to measure the dust kinematics and compare with other phases of the ISM. All the Seyfert galaxies host both AGN-driven outflows and ongoing star formation which makes them ideal targets to study the potential dust phase of AGN outflows and their impact on their host galaxy. The paper is laid out as follows. In Section \ref{sec:Data}, we describe the sample and the observations with JWST. In Section \ref{sec:Methods}, we describe the method of PCA tomography. In Section \ref{sec:Results} we present velocity maps for each target. Finally, in Section \ref{sec:Discussion} we discuss our results. Throughout this work we assume $\Lambda$CDM cosmology with $H_0 = 70$ km s$^{-1}$ Mpc$^{-1}$, $\Omega_m = 0.27$, $\Omega_{\Lambda} = 0.73$.


\section{Observations and Data Reduction}
\label{sec:Data}
In this work we use 10 local Seyfert galaxies as part of the GATOS (Galaxy Activity Torus and Outflow Survey) sample observed with both MIRI MRS and NIRSpec IFU providing spatially resolved spectroscopy from $\sim 3\mu$m - $28\mu$m. A summary of the physical properties of these targets are shown in Table \ref{tab:Sample}. The data are compiled from  four JWST General Observer programs over cycles 1, 2, and 3. These are GO 1670 (P.I. T. Shimizu), GO 3535 (P.I. I. Garcia-Bernete $\&$ D. Rigopoulou), GO 5017 (P.I. I. Garcia-Bernete $\&$ A. Alonso-Herrero $\&$ D. Rigopoulou), and GO 1875 (P.I. J. Fernandez Ontiveros). 

All of the targets contain outflows clearly detected in the ionised phase \citep[e.g.][]{Davies2020} as well as star-formation in the circumnuclear regions which have strong PAH emission, although some targets are much more dominated by the nuclear emission such as MCG-05-23-016 \citep[e.g.][]{Gonzalez-Martin2025, Esparza-Arredondo2025}.



\begin{table*}
\centering
  \caption{Summary of Targets}
  \label{tab:Sample}
    \def\arraystretch{1.7}
    \setlength{\tabcolsep}{3pt}

    \begin{threeparttable}
  \begin{tabular}{ccccccccccc}
  
    \hline

    Name & RA & Dec & $z$ &  Observing Program & $\log L_{\rm bol}$ & $\log\lambda_{\rm Edd} $ & $\log M_{\rm BH}$& Disk PA & Outflow PA & Ref.\\
     & & & & & (erg s$^{-1}$) & & ($M_{\odot}$) & & &\\
    \hline
  ESO 420-G013 &  04h13m49.6804s & -32d00m25.153s	& 0.012 & GO 5017, GO 1875 & - & -&- &230$^{\circ}$ & 32$^{\circ}$ & (1) \\
MCG-05-23-016 & 09h34m06.3120s & +27d20m59.402s	& 0.042 &  GO 5017, GO 1670 & 44.3 & -1.70& 7.9 &246$^{\circ}$  & 172$^{\circ}$ & (2),(3),(4),(5)\\
NGC 3081 & 09h59m29.5431s & -22d49m34.720s	 & 0.008 &  GO 1670, GO 5017 & 44.1 &-1.20 & 7.20 & 90$^{\circ}$ & 180$^{\circ}$ & (6),(4),(7) \\
NGC 3227 & 10h23m30.5748s & +19d51m54.340s & 0.004 & 
GO 3535 & 43.3 &-1.84 &7.04 & 332$^{\circ}$ & 30$^{\circ}$ & (8),(9),(10),(11)\\
NGC 4051 & 12h03m09.6073s & +44d31m52.709s & 0.002 & 
GO 3535 & 42.4 & -1.94 & 6.24 & 50$^{\circ}$ & 80$^{\circ}$ & (9),(12) \\
NGC 5506 & 14h13m14.8756s & -03d12m27.698s & 0.006 &  GO 1670 & 44.1 &-1.30& 7.3 &271$^{\circ}$ & 22$^{\circ}$ & (9),(13),(4) \\
NGC 5728 & 14h42m23.8735s & -17d15m10.855s & 0.009 &  GO 5017, GO 1670 & 44.1 & -1.53 & 7.53 & 14$^{\circ}$ & 293$^{\circ}$ & (14),(15),(16) \\
ESO 137-G034 & 16h35m14.0026s & -58d04m47.869s & 0.009 & GO 1670, GO 5017 & 43.4 &-2.10 & 7.4 &348$^{\circ}$ & 304$^{\circ}$ & (3),(4),(17) \\
NGC 7172 & 22h02m01.8945s & -31d52m10.525s & 0.009 &  GO 1670& 44.1 & -2.30 & 8.3 &90$^{\circ}$ & 0$^{\circ}$ & (18),(4),(19)\\
NGC 7582 & 23h18m23.6280s & -42d22m13.512s & 0.005 &  GO 3535 & 43.3 & -1.42 & 6.62 &168$^{\circ}$ & 221$^{\circ}$ & (20),(21)\\

    \hline
  
  \end{tabular}
\begin{tablenotes}
\item[] Observing Programs: GO 1670 (P.I. T. Shimizu), GO 3535 (P.I. I. Garcia-Bernete $\&$ D. Rigopoulou), GO 5017 (P.I. I. Garcia-Bernete, A. Alonso-Herrero, D. Rigopoulou), GO 1875 (P.I. J. Fernandez Ontiveros). 
\item[] Both the disk and outflow position angles are measured east of north to the blueshifted side. 
\item[] References numbered by first appearance in the table:
(1) \citep{Fernandez-Ontivero2020}  
(2) \citep{Esparza-Arredondo2025}  
(3) \citep{Zhang2024b}  
(4) \citep{Davies2020}  
(5) \citep{Ponti2012}  
(6) \citep{Schnorr-Muller2016}  
(7) \citep{Gliozzi2024}  
(8) \citep{Alonso-Herrero2019}  
(9) \citep{Fischer2013}  
(10) \citep{Bentz2023}  
(11) \citep{Mehdipour2021}  
(12) \citep{Denney2009}  
(13) \citep{Gofford2015}  
(14) \citep{Shimizu2019}  
(15) \citep{Davies2015}  
(16) \citep{Durre2018}  
(17) \citep{Caglar2020}  
(18) \citep{Alonso-Herrero2023}  
(19) \citep{Smajic2012}  
(20) \citep{Veenema2026}  
(21) \citep{Poitevineau2025}
\end{tablenotes}
  \end{threeparttable}
 \end{table*}




We primarily followed the standard pipeline procedure (e.g., \citealt{Labiano2016} and references therein) and the same configuration of the pipeline stages described in \citet{Garcia-Bernete2022c} and \citet{Pereira-Santaella2022} to reduce the data. Some hot and cold pixels are not identified by the current pipeline, so we added some extra steps, as is described in \citet{Pereira-Santaella2024} and \citet{Garcia-Bernete2024b} for NIRSpec and MRS, respectively. We used calibration version 1.18.1 with CDRS version 12.1.4 for the data reduction.

The NIRSpec IFU observations were done using the high spectral resolution mode with $R\sim2700$ in the g395h-f290lp filter, with a 4 point dither to properly sample the PSF. This gives a wavelength range of 2.87 $\mu$m - 5.27 $\mu$m. From the NIRSpec data, the only feature we produce velocity maps for is the 3.3 $\mu$m PAH feature as presented in section \ref{sec:Results}.

The MIRI MRS cubes provides data from $\sim5\mu$m-$28\mu$m, with the field of view increasing but the spatial resolution decreasing with wavelength. From the data reduction process, 12 cubes are produced, splitting up the wavelength range by channels, 1, 2, 3, 4, which each have sub channels A, B, C. For this work we do not use any features in channel 4, with our longest wavelength feature, H$_2$ S(1) line (17.03 $\mu$m) appearing in the channel 3-long cube. Therefore we have an R $\sim3500 - 2000$, from $\sim5\mu$m - $18\mu$m respectively.

Both the MIRI and NIRSpec cubes are produced in the IFU alignment, where the spatial axes are aligned with the position angle of the telescope when the observations were taken. Any analysis is performed on the cubes in the IFU alignment before being rotated into the sky orientation (north is up, east is left) only when plotting. This reduces any further errors that may be introduced when reprojecting the data.

\section{PCA Tomography}
\label{sec:Methods}

\begin{figure*}
\hspace*{0.75cm}                                           
	\includegraphics[width=14cm]{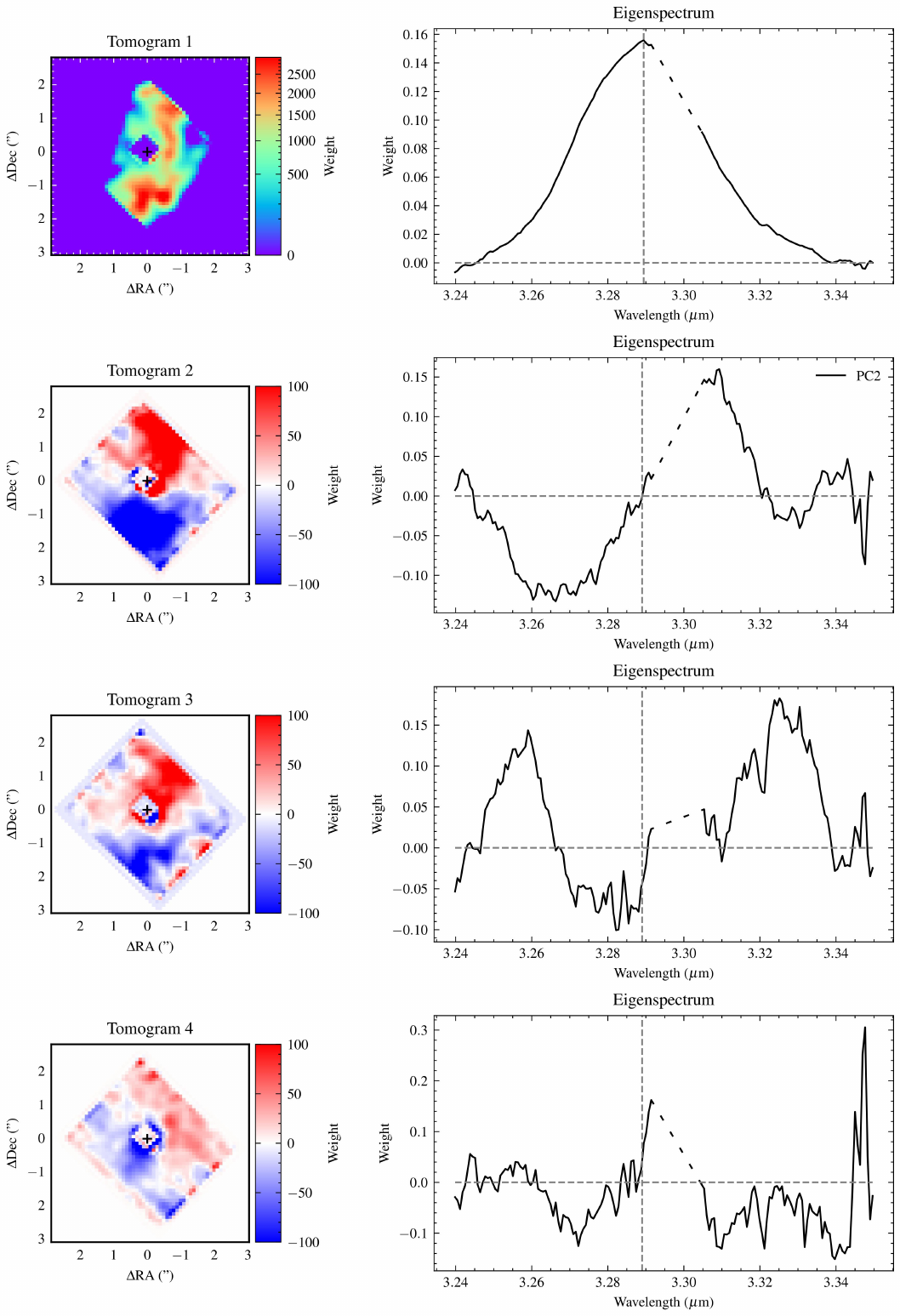}
    \caption{The first four principal components of the 3.3 $\mu$m PAH for NGC 7582. Each row shows each component, where the higher components contain progressively less information held within the original data cube. The left panels show the tomograms which show the spatial distribution of the eigenspectra in the right panels. The first component describes the rest frame emission of the feature while the second component shows a redshifted/blueshifted feature on either side of the nucleus, indicative of rotation. The higher order components contain extra information on the velocity field and/or variations in the intrinsic profile of the PAH feature. The dashed section of the eigenspectra are where P$\alpha$ has been masked from the data. Note that the colour scale for each of the tomograms is not velocity but rather a correlation coefficient of how strongly each eigenspectrum maps onto the spatial dimensions. The vertical dashed grey lines show the wavelength where the second eigenspectrum crosses zero, showing the kinematic centre. Note that the smooth line between 3.29 - 3.30 $\mu$m that can be seen in the Eigenspectra is due to the masking of the Pf$\delta$ line.}
    \label{fig:PCADemo} 
\end{figure*}

The technique of PCA tomography was first presented in \citet{Steiner2009} and applied to JWST data in \citet{Donnan24b} to reveal the kinematics of PAH features for the first time. We summarise the technique below before demonstrating the technique in Fig. \ref{fig:PCADemo}.

The purpose of PCA is to reorder the data into principal components which contain a decreasing fraction of the total information held within the data. These components are linear, meaning the original data can be reconstructed by summing the principal components. When applied to data cubes containing two spatial dimensions and one spectral dimension, each principal component contains a specific spectral feature that has some distribution in the spatial dimensions. 

Before applying the PCA decomposition, the cubes are prepared in the following way. First the spectral feature of interest is isolated in wavelength, creating a sub-cube of only the a single emission feature to apply the PCA to. The exact wavelength limits depend on how broad the feature is, with the PAHs having a larger wavelength window than any of the emission lines. For example, for the 3.3 $\mu$m PAH we use data only between 3.24 $\mu$m and 3.35 $\mu$m while for the H$_2$ S(5) we use a window between 6.87 $\mu$m and 6.95 $\mu$m.

By isolating the emission feature in its own sub-cube, we ensure that the number of spectral dimensions is less than the total number of spatial pixels which is essential for PCA to successfully decompose the data. This is because the the spatial axes contain the information that is used to separate the principal components and avoid the ``curse of dimensionality'' \citep[][]{Takeuchi2024}. For example, in Fig. \ref{fig:PCADemo}, we isolate the 3.3 $\mu$m PAH feature resulting in spatial dimensions of 45$\times$39 creating a total of 1730 spatial pixels (after subtracting 25 pixels for the masked PSF at the nucleus in this specific example) while there are only 167 spectral pixels. Moreover, the angular and spectral resolution of the data is approximately constant over the sub cube due of a given spectral feature.

A local continuum is then fit assuming a simple straight line before it being subtracted. The resulting cube has the form $\mathbf{I}_{x, y, \lambda}$ which is then collapsed into two dimensions where the spatial dimensions, $x, y$ are transformed by
\begin{equation}
\label{eqn:CoordTrans}
    \beta = \mu (x-1) +y.
\end{equation}
resulting in a cube,  $\mathbf{I}_{\beta, \lambda}$. The PCA is applied to this cube where 
\begin{equation}
    \mathbf{T}_{\beta, k} = \mathbf{I}_{\beta, \lambda} \cdot \mathbf{E}_{\lambda, k}, 
\end{equation}
where $\mathbf{T}_{\beta, k}$ are the tomograms of each principal component, $k$, which describes the spatial distribution of a given spectral feature, $\mathbf{E}_{\lambda, k}$, which are known as eigenspectra. We additionally mask low signal to noise spaxels as this can ``confuse'' the PCA as it treats each spaxel with equal weight. We measure the signal to noise by first measuring the noise by smoothing the spectrum of each spaxel (in the case of a PAH feature) and subtracting from the data. The noise is then measured as the standard deviation of the residuals. In the case of an emission line, we mask the line and measure the standard deviation. The signal to noise is then the peak flux of the feature over the noise. We mask spaxels with a signal to noise ratio < 5. We additionally smooth each channel of the data cube with a Gaussian kernel with $\sigma = 1$ pixel to improve the signal to noise before masking. For the 17 $\mu$m PAH, we increase the smoothing to 2 pixels as well as lowering the signal to noise mask to <1, as the data is significantly noisier for this particular feature. The 17 $\mu$m PAH is comprised of several sub-bands however we only use the 16.4 $\mu$m sub feature to measure the kinematics of the 17 $\mu$m PAH, as the full feature is extremely broad while the 16.4 $\mu$m sub feature is relatively narrow.

Additionally for the 3.3 $\mu$m PAH and the 6.2 $\mu$m PAH, we mask emission lines that are blended with the feature. In particular at 3.296 $\mu$m the HI 9-5 line is masked as well as the H$_2$ S(6) line at 6.109 $\mu$m. 

We demonstrate the PCA technique applied to the 3.3 $\mu$m PAH for NGC 7582 in Fig. \ref{fig:PCADemo}, where we show the first four principal components. From the PCA decomposition it is clear that the first principal component describes the rest frame emission of the feature, where the eigenspectrum shows the restframe profile while the tomogram shows the emission map The second principal component contains kinematic information, where the eigenspectrum shows a lack of emission on the blue side but an increase in emission on the red side, corresponding to a redshifted profile. Where the second tomogram is positive, the feature is redshifted and where the tomogram is negative the feature is blueshifted. 

\begin{figure}
	\includegraphics[width=\columnwidth]{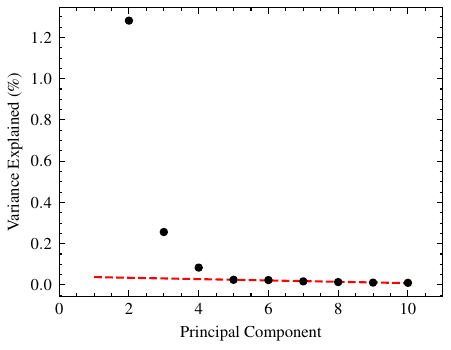}
    \caption{Scree test of the PCA decomposition of the 3.3 $\mu$m PAH for NGC 7582 (as shown in Fig. \ref{fig:PCADemo}) which shows the percentage of variance that each principal component explains in the data. The first principal component contains 98$\%$ of the variance and so it appears of the scale. The variance explained levels off after the 4\textsuperscript{th} component where no useful information is contained and is shown with the dashed red line. }
    \label{fig:Scree} 
\end{figure}

The higher order components contain progressively less information, eventually only containing noise. This is demonstrated by a Scree test \citep[][]{Cattell1996} which is plotted in Fig. \ref{fig:Scree} and demonstrates the variance of the original data held in each principal component. After the fourth component, the plot levels off indicating no significant information is held in these components, with only $\sim0.1\%$ of the variance held in the sum of the remaining components.

In Fig. \ref{fig:PCADemo}, the third and fourth components may contain additional velocity information and/or small differences in the intrinsic profile. Considering the third tomogram is similar to the second, it likely contains additional velocity information, however the fourth tomogram is not particularly clear and may just be noise. As we will show in section \ref{sec:PAHProp}, in the case 6.2 $\mu$m PAH, small differences in the intrinsic profile are present instead of kinematics. The tomogram is powerful for distinguishing between kinematics and variations in the intrinsic profile as it can be compared to known line kinematics. When variations in the intrinsic profile do show up, it is clear from the tomogram, which is indeed the case for the 6.2 $\mu$m PAH which we discuss in section \ref{sec:PAHProp}. We find that the 3.3 $\mu$m, 11.3 $\mu$m and 17 $\mu$m PAHs all show kinematics rather than differences in the intrinsic profile.

To construct a velocity map from the principal components, we take the first principal component as the rest frame emission over the field of view. The eigenspectrum for this component is then used as a reference to measure the velocity for each spaxel. We do this by reconstructing the original cube using the first 4 components (subsequent components contain no useful information as demonstrated by the Scree test in Fig. \ref{fig:Scree}) and then measuring the velocity offset between the spectrum of a given spaxel and the eigenspectrum of the first principal component. The velocity of a spaxel is given by
\begin{equation}
    V_{x, y} = 3\times10^5 \Delta\lambda_{x, y}/\lambda_0 \,\, \mathrm{km\,s}^{-1} ,
\end{equation}
where $\Delta\lambda_{x, y}$ is the shift measured between the first principal component and each spaxel of the reconstructed data cube using the first 4 principal components. $\lambda_0$ is the reference wavelength, which in the case on an emission line is just the rest wavelength of the feature. For the PAH features, there is no clear reference wavelength due to the intrinsic broadness and asymmetry of the profile. We therefore use the wavelength where the second principal component (which contains the majority of the kinematic information) crosses from positive to negative as demonstrated in Fig. \ref{fig:PCADemo}. This point represents the kinematic centre, making it a good measure of $\lambda_0$.

\citet{Donnan24b} demonstrated that this technique is effective in recovering the velocity field as compared to more traditional methods. The necessity to use PCA tomography to measure the velocity field of PAH features over more traditional methods such as fitting each spaxel, results from two major advantages it has over simple velocity cross-correlation. First, the PCA can detect shifts in the restframe feature, using all the information contained within the spatial dimensions, picking out the signature of kinematics.  Secondly, there is not well defined a priori restframe profile for PAH features and so the first principal component provides a reference profile to be able to measure the velocity field.  

We did some further tests to ensure that the PCA technique can accurately infer velocities. First we can look at the shape of the eigenspectrum of the second principal component which should match the derivative of the eigenspectrum of the first principal component if a Doppler shift is present. This test is presented in Appendix \ref{sec:IntProf}.

\section{Results}
\label{sec:Results}
\subsection{Velocity Maps}
We run the PCA analysis for each target for a variety of spectral features, probing different phases of the ISM. PCA is used for to produce all of the velocity maps we present in this work. We confirm that the PCA technique is robust by checking that we get the same velocity maps through a more traditional method of modelling the spectrum of each spaxel with a Gaussian to infer a velocity. This can be found in Appendix \ref{sec:LineTest}. 

To clearly isolate the outflow we use the [NeVI] (7.65 $\mu$m), which has an ionisation potential of 158 eV making it an unambiguous tracer of AGN photoionisation \citep[e.g.][]{Sturm2002}. 

We then probe the molecular gas using the rotational transitions of H$_2$, where we use the S(5) (6.91 $\mu$m) to probe the hot molecular gas and the S(1) (17.03 $\mu$m) to probe the cooler molecular gas. As \citet{Davies2024} demonstrated, if the outflow is detected in the molecular phase, the hotter H$_2$ transitions trace more of the outflow than the cooler transitions, where CO (2-1), probing even cooler molecular gas, shows no detection in the outflow \citep[][]{Shimizu2019}. It is worth noting however that the relatively low FOV of MIRI means we may miss molecular outflows on very large scales ($\gtrsim$ kpc).


We extract the kinematics of the PAH features, providing a probe of the dust phase and in particular small carbonaceous dust. We focus on the 3.3$\mu$m, 11.3 $\mu$m and 17.0 $\mu$m (via the 16.4 $\mu$m sub-feature) features due to their relative lack of blending with other features. 
We note that the results using the 6.2 $\mu$m PAH feature do not easily compare with what we see from the other features. This might be because the 6.2 $\mu$m PAH feature arises from ionised PAHs which could be subject to destruction by the radiation field of the AGN \citep[e.g.][]{Garcia-Bernete2024b}, leading to a relatively low signal to noise in outflow regions compared to other PAH features. The PCA appears to select different intrinsic profiles for the outflow regions compared to SF regions for the 6.2 $\mu$m feature. We therefore discuss the output of the PCA analysis of the 6.2 $\mu$m PAH in section \ref{sec:PAHProp}. 
We also avoid the 7.7 $\mu$m due to the blending with many emission lines such as [NeVI] (7.65 $\mu$m) and the complex continuum in this region of the mid-infrared. Therefore our analysis of the 3.3, 11.3 and 17.0 $\mu$m features are strongly dominated by the neutral PAHs.

We expand on earlier work of \citet{Donnan24b} by including the longer wavelength PAH features. These provide challenges as the 11.3 $\mu$m and 17.0 $\mu$m PAHs are broader than the 3.3 $\mu$m PAH and also have a poorer spatial resolution. However the larger field of view compensates for this loss, where the greater number of spatial pixels results in more information available for the PCA decomposition. By probing different PAH bands, we are sensitive to different PAH molecules. The 3.3 $\mu$m PAH originates mainly from small grains while the 11.3 $\mu$m feature is dominated by larger, neutral grains \citep[e.g.][]{Rigopoulou2021}. The 6.2 $\mu$m PAH mainly originates from ionised PAHs.

As described in section \ref{sec:Methods}, velocity maps can be recovered from the principal components, where the first component contains the restframe emission while the higher order components and in particular the second component, contain the information about the velocity field. We show the resulting velocity maps for all the features described above in Fig. \ref{fig:VelMaps}.


\begin{figure*}
    \centering
        \includegraphics[trim={0 40.7cm 0 0},clip,width=\textwidth]{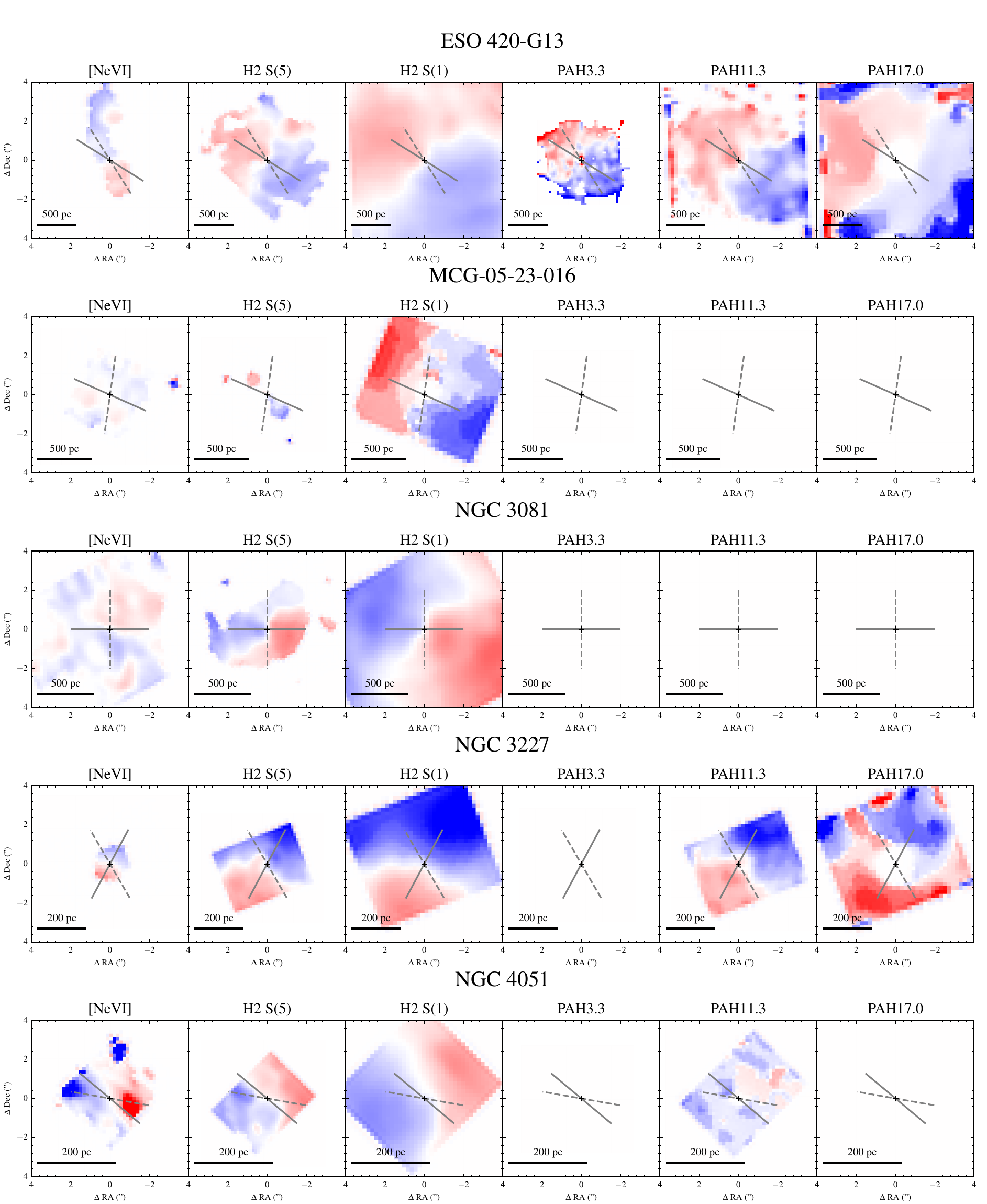}
    \phantomcaption
\end{figure*}

\begin{figure*}
\ContinuedFloat
    \centering
        \includegraphics[trim={0 0 0 38.95cm},clip,width=\textwidth]{Images/Test.pdf}
    \caption{Velocity maps of each spectral feature for each galaxy, all of which are constructed using PCA tomography. From left to right we show [NeVI] (7.65 $\mu$m), H$_2$ S(5) (6.91 $\mu$m), H$_2$ S(1) (17.03 $\mu$m),  3.3$\mu$m PAH, 11.3 $\mu$m PAH and the 17 $\mu$m PAH. For each galaxy the field of view is the same for each map as well as the velocity scale. We show [NeV] (14.32 $\mu$m) for NGC 7582 as the [NeVI] (7.65 $\mu$m) has a lower signal to noise. The position angle of the major axis of the disk is shown with the solid line while the position angle of the major axis for the outflow is shown with the dashed line. Values are found in Table \ref{tab:Sample}. Panels that are blank lack the signal to noise to produce velocity maps.}
    \label{fig:VelMaps}
\end{figure*}

\subsubsection{Ionised Gas}
We use the [NeVI] (7.65 $\mu$m, 158 eV) emission line to trace highly ionised gas due to AGN photoionisation \citep[e.g.][]{Zhang2024b}. Some of the targets, such as NGC 5506 and NGC 7172 are edge-on galaxies, where the ionisation cone extends to the north and south of the disk \citep[e.g.][]{HermosaMunoz2024} and therefore there very little of the velocity field is along the line of sight making the outflow kinematics difficult to detect. We also find MCG-05-23-016 to provide a weak detection, where the emission is point-like and associated with the nucleus. 

For NGC 7582 we opt to use [NeV] (14.32 $\mu$m 77.4 eV) instead of [NeVI] as the signal to noise is higher making the kinematics of the outflow much clearer \citep[][]{Veenema2026}.


\subsubsection{Molecular Gas}

The hot molecular gas is traced by the H$_2$ S(5) line while cooler molecular gas is traced by the H$_2$ S(1) line. The H$_2$ kinematics are typically dominated by rotation in the galaxy disk, however the S(5) line appears to show contribution from the outflow in ESO 137-G034, NGC 5728 and NGC 7582. This result was presented by \citet{Davies2024} for NGC 5728 while \citet{Veenema2025} presents analysis of NGC 7582.

The increase in the contribution of the outflow results in the position angle of the kinematic axis changing from the S(1) to the S(5) line. For NGC 7582, the picture is less clear for the H$_2$ lines. This is partly due to the asymmetric FOV of the hotter H$_2$ lines, making it difficult to fully observe the kinematics. Moreover, the flux of the H$_2$ lines is strongly dominated by two star-forming regions to the south east \citep[][]{Veenema2025}, which also show excess blueshifted velocities in the molecular gas \citep[][]{Veenema2025, Garcia-Burillo2021}. However we do see some excess redshifted emission towards the north east consistent with the velocity map of [NeV] which may suggest a small outflow contribution in the H$_2$ lines for NGC 7582. This is investigated further in section \ref{sec:DiskModelling}. Alternatively, \citet{Garcia-Burillo2021} suggests that the redshifted excess may result from the projection of local inflowing gas considering they coincide with the passage of spiral arms along the minor axis.

As with the ionised gas, the edge-on nature of NGC 5506 and NGC 7172 makes it challenging to detect their outflows kinematically. The remaining objects, ESO 420-G13, MGC-05-23-16, NGC 3081, NGC 3227 and NCG 4051 do not show any detectable contribution to the outflow in the molecular gas as traced by the rotational transitions of H$_2$ \citep[][]{Esparza-Arredondo2025}.

\subsubsection{PAHs}

In objects where the signal-to-noise of the 3.3 $\mu$m PAH is high enough to be detected, the velocity is consistent with rotation in the circumnuclear disk and shows no detectable outflow contribution in any of the objects. However the 11.3 $\mu$m and 17.0 $\mu$m PAHs are more consistent with the rotational H$_2$ transitions than the ionized gas and show contribution from the outflow in two objects: NGC 5728 and NGC 7582. We see no evidence for the outflow in the PAH velocity maps of NGC 3227, NGC 4051 and ESO 420-G13. The remaining objects, MCG-05-23-016, NGC 3081, NGC 5506, ESO 137-G034 and NGC 7172 either lack the signal to noise or appear edge-on, meaning we cannot rule out the presence of PAHs in their outflows, but currently are hard to detect.

In NGC 5728, the 11.3 PAH shows excess redshifted emission in the redshifted side of the outflow. In NGC 7582 the outflow is even clearer, with the 11.3 $\mu$m PAH velocity map dominated by the outflow, consistent with the hot molecular gas as traced by the H$_2$ S(5) line. This is the first spatially resolved kinematic evidence of dust in the outflow of an AGN. 

For NGC 5728, the 11.3 $\mu$m PAH shows rotation consistent with circumnuclear disk but also clear redshifted emission at the base of the redshifted side of the outflow. However the blueshifted side of the outflow does not appear to be present in the 11.3 $\mu$m velocity map. As this part of the outflow is behind and obscured by the star-forming ring \citep[e.g.][]{Davies2024}, the velocity signature may be difficult to detect. Unlike the H$_2$ lines or ionised gas where the outflow is bright in the flux (moment 0) map, the PAHs are significantly brighter in the star-forming ring than in the outflow. This means the velocity (moment 1) signal of the outflow within the data is much less significant (due to the flux differences) than the disk in this region for the PAH features, while the H$_2$ lines and ionised lines have bright flux in the outflow, making the blueshifted velocity signal much easier to detect despite the obscuration by the star-forming ring. 

Considering the 9.8 $\mu$m silicate absorption feature can affect the shape of the 11.3 $\mu$m PAH feature, we checked to see if extinction can explain the redshifted emission we see. Due to the silicate feature, if the extinction changes drastically spatially, the PAH feature will be suppressed more at shorter wavelengths, mimicking a redshift. However, we find that the extinction is relatively low in the outflow regions of NGC 5728, with low silicate depths. Therefore extinction is likely not significantly affecting the PAH emission in the outflow of NGC 5728.

The velocity map of the 17.0 $\mu$m PAH is noisier, however it does show a clear warp in the position angle of the kinematic axis, where there is excess redshifted emission due to the outflow. We investigate the kinematic signature of the outflow in the PAH velocity maps further in section \ref{sec:DiskModelling} where we isolate the outflow and disk components. 

The velocity maps of the 11.3 $\mu$m and 17.0 $\mu$m PAHs for NGC 7582 show clear contribution from the outflow, where the redshifted emission to the north east is from the outflow whereas the redshifted emission to the north west is from the disk. In-fact, the 11.3 $\mu$m PAH appears to be dominated by the outflow, closer matching the [NeV] velocity map rather than the 3.3 $\mu$m PAH velocity map. The 3.3 $\mu$m PAH only shows the kinematics of the disk, where only the redshifted emission to the north west is present. We also investigate this object further in section \ref{sec:DiskModelling} by modelling the disk kinematics.

\subsection{Disk modelling}
\label{sec:DiskModelling}
As the signal to noise of NGC 7582 and NGC 5782 is sufficiently high and show potential evidence of outflow in the kinematics of the 11.3 $\mu$m PAH, we opt to model and subtract the disk kinematics to reveal the outflow kinematics. This was demonstrated in \citet{Davies2024}, where the velocity maps of H$_2$ contain contribution from both the circumnuclear disk and the
outflow. To recover the outflow component, one can model the velocity field of the disk and subtract from the observed velocity map of a given feature. 

We follow \citet{Veenema2026} and fit a disk model at a fixed inclination and position angle constrained from previous observations. The free parameters of the model are the parameters governing the intrinsic radial velocity curve, $ V(R)$, which has the form of a $\tanh$ function allowing the rotation curve to rise steeply at low radii and then flatten at larger radii,
\begin{equation}
    V(R) = V_{\textrm{max}}\tanh{\left(\frac{R}{R_{\textrm{turn}}}\right)},
\end{equation}
where the $V_{\textrm{max}}$ and $R_{\textrm{turn}}$ are free parameters. The observed velocity, $V_{\rm obs}(x,y)$, of each spaxel $(x,y)$ is therefore 
\begin{equation}
    V_{\rm obs}(x,y) = V_{\rm sys} + V(R)\,\sin i \,\cos\theta,
\end{equation}
where $V(R)$ is the velocity of each ring at radius $
R$ and $V_{\rm sys}$ is the systemic velocity. The radius, $R$, is the intrinsic, deprojected radius which relates to the coordinates of each spaxel $(x,y)$ via
\[
\begin{aligned}
x' &= x\cos\phi + y\sin\phi, \\
y' &= -x\sin\phi + y\cos\phi, \\
R &= \sqrt{x'^2 + \left(\frac{y'}{\cos i}\right)^{2}}, \\
\cos\theta &= \frac{x'}{R}, \\
\end{aligned}
\]
where $\phi$ is the position angle of the disk on the sky.


Typically one can fit a velocity model, such as the tilted ring model, to a 3D data cube using codes such as 3D-Barolo \citep[][]{DiTeodoro2015}, modelling both the moment 0 and moment 1 maps of a given feature. However we fit to the 2D velocity maps (moment 1) of a given feature instead as the PAH features, unlike emission lines, have an intrinsic broad profile. We fit each of the velocity maps for every emission feature in NGC 5728 and NGC 7582, using MCMC sampling from \textsc{NUMPYRO} \citep{Phan2019}. 

\begin{figure*}
	\includegraphics[width=\textwidth]{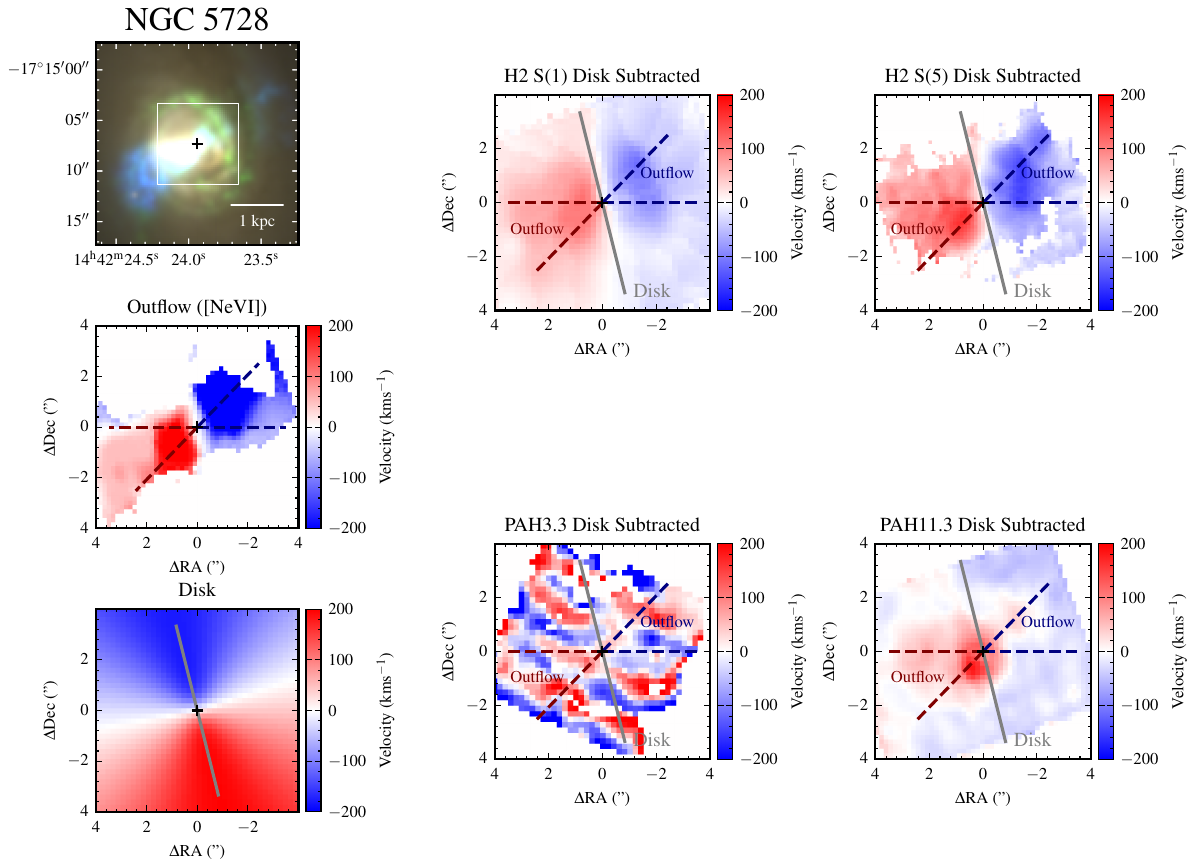}
    \caption{Disk modelling and subtraction for NGC 5728. The top left panel shows a three colour image constructed from custom filters of MUSE cubes. Blue is a filter centred on [O III] (5028–5095 {\AA}), green is H$\alpha$ (6578–6664 {\AA}) and an $i$ band filter (6430–8630 {\AA}) is shown in red. The white square shows the FOV of the other panels. The middle left panel shows the velocity map of [NeVI] tracing the outflow. The bottom left panel shows the disk model with fixed inclination and position angle constrained from CO and stellar kinematics \citep[][]{Shimizu2019}. The right panels show the disk subtracted velocity maps for molecular gas (warm S(1) and hot S(5)) as well as the 3.3 $\mu$m PAH (small $\&$ neutral) and the 11.3 $\mu$m PAH (large $\&$ neutral). The major axis of the disk is shown with the solid grey line while the outflow cone is shown with the dashed lines in red and blue for the redshift and blueshift sides of the cone respectively.}
    \label{fig:DiskModellingNGC5728} 
\end{figure*}

For NGC 5728, the geometry of the disk is well constrained from both the CO (2-1) velocity field and the stellar kinematics \citep[][]{Shimizu2019}. We found that directly using the disk models presented in \citet{Shimizu2019}, over-predicted the rotation velocity for the 11.3 $\mu$m PAH feature. We therefore fix the position angle/inclination contained by the stellar/CO kinematics of PA$=14^{\circ}$, $i =43 ^{\circ}$ and fit for the rotational velocity. We present the results of the disk modelling for NGC 5728 in Fig. \ref{fig:DiskModellingNGC5728}.

For NGC 7582 we use the position angle constrained from the low ionisation potential lines in the MIRI data from \citet{Veenema2026} of PA=$221^{\circ}$. We use an inclination of $i=58^{\circ}$ \citep[][]{Morris1985, Wold2006} consistent with \citet{Veenema2026} and the stellar kinematics presented in \citet{Juneau2022}. We fix the inclination and position angle of the disk in our fitting using the above values.

\begin{figure*}
	\includegraphics[width=\textwidth]{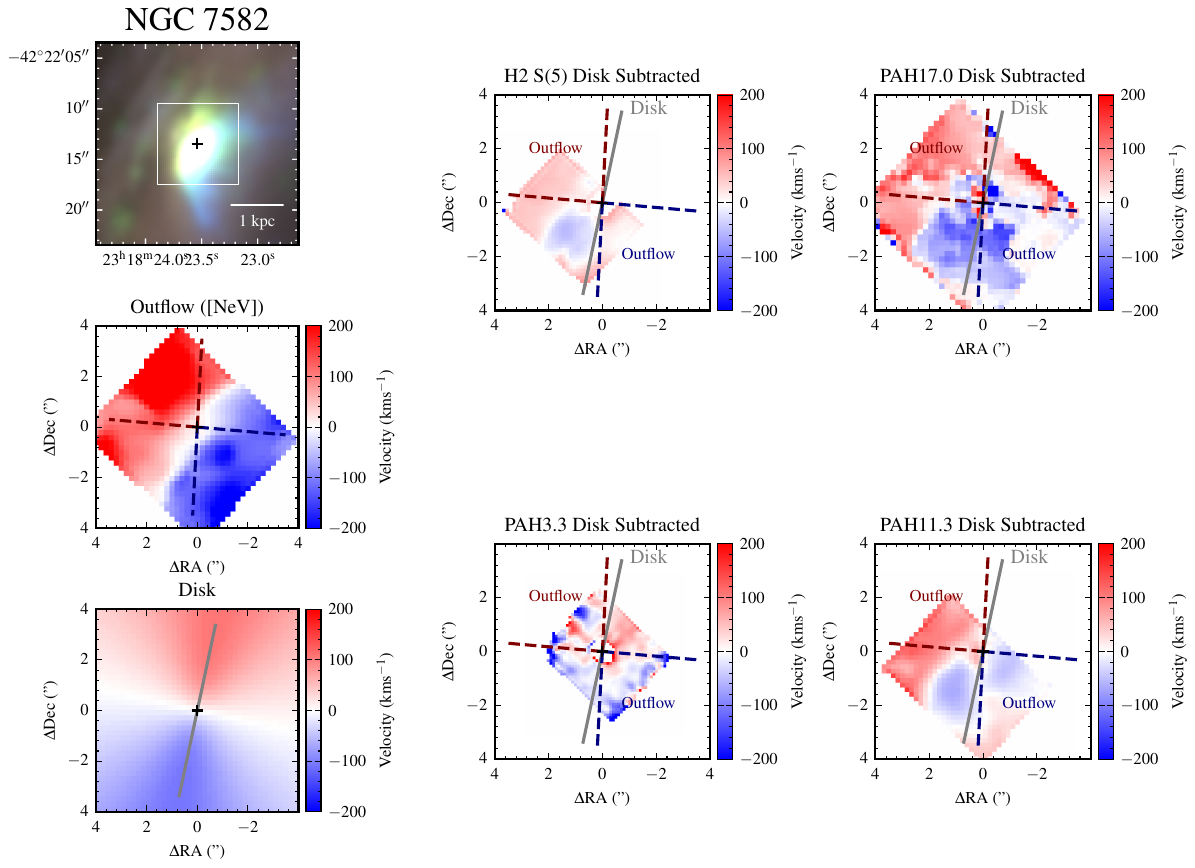}
    \caption{Disk modelling and subtraction for NGC 7582. The top left panel shows a three color image constructed from custom filters of MUSE cubes. Blue is a filter centred on [O III] (5028–5095 {\AA}), green is H$\alpha$ (6578–6664 {\AA}) and an $i$ band filter (6430–8630 {\AA}) is shown in red. The white square shows the FOV of the other panels. The middle left panel shows the velocity map of [NeV] tracing the outflow. The bottom left panel shows the disk model with fixed inclination and position angle. The right panels show the disk subtracted velocity maps for molecular gas (S(5)) as well as the 3.3 $\mu$m PAH (small $\&$ neutral), the 11.3 $\mu$m PAH (large $\&$ neutral) and the 17$\mu$m PAH (large $\&$ neutral). The major axis of the disk is shown with the solid grey line while the outflow cone is shown with the dashed lines in red and blue for the redshift and blueshift sides of the cone respectively.}
    \label{fig:DiskModellingNGC7582} 
\end{figure*}

This results in the disk models shown in Fig. \ref{fig:DiskModellingNGC7582}. Subtracting the disk model from the observed velocity map results in the right hand column of Fig. \ref{fig:DiskModellingNGC7582}. We additionally shift the systemic velocity of the H$_2$ S(5), H$_2$ S(1) and the 11.3 $\mu$m PAH, by adding the systemic velocity as a free parameter to the disk fitting, as the maps appear to be dominated by redshifted emission. This is likely a result of the field of view being off-centre, where the emission is therefore dominated by the blueshifted side of the disk. This leads to the first principal component appearing too blue and therefore the wavelengths shifts for each spaxel are redder resulting in a map offset to positive velocities. 

We find that the 3.3 $\mu$m PAH is fit well by the disk model, with the residuals not showing any clear outflow signature. The opposite picture is true however for the 11.3 $\mu$m PAH, where the position angle of the velocity map is much more consistent with [NeV] than the disk kinematics \citep[][]{Juneau2022}, even without modelling the disk component.  The 17 $\mu$m PAH also shows excess redshifted and blueshifted velocities consistent with [NeV] but has a stronger disk component than the 11.3 $\mu$m PAH. 

As mentioned previously, the outflow is not as clear in the H$_2$ lines for NGC 7582. The residuals of the H$_2$ S(5) line do show residual redshifted gas consistent with the outflow however there is also an excess of blueshifted gas to the south east, coincident with H$_2$ bright star-forming regions \citep[][]{Veenema2025}. From CO lines, \citet{Garcia-Burillo2021} points out that the excess redshifted and blueshifted velocities in the H$_2$ lines could be the projection of local inflows rather than outflows, considering the coincidence with spiral arms along the minor axis of the disk. However they also do note that there are coplanar outflow motions in the cold H$_2$ as well albeit on small scales.


\begin{figure*}
	\includegraphics[width=\textwidth]{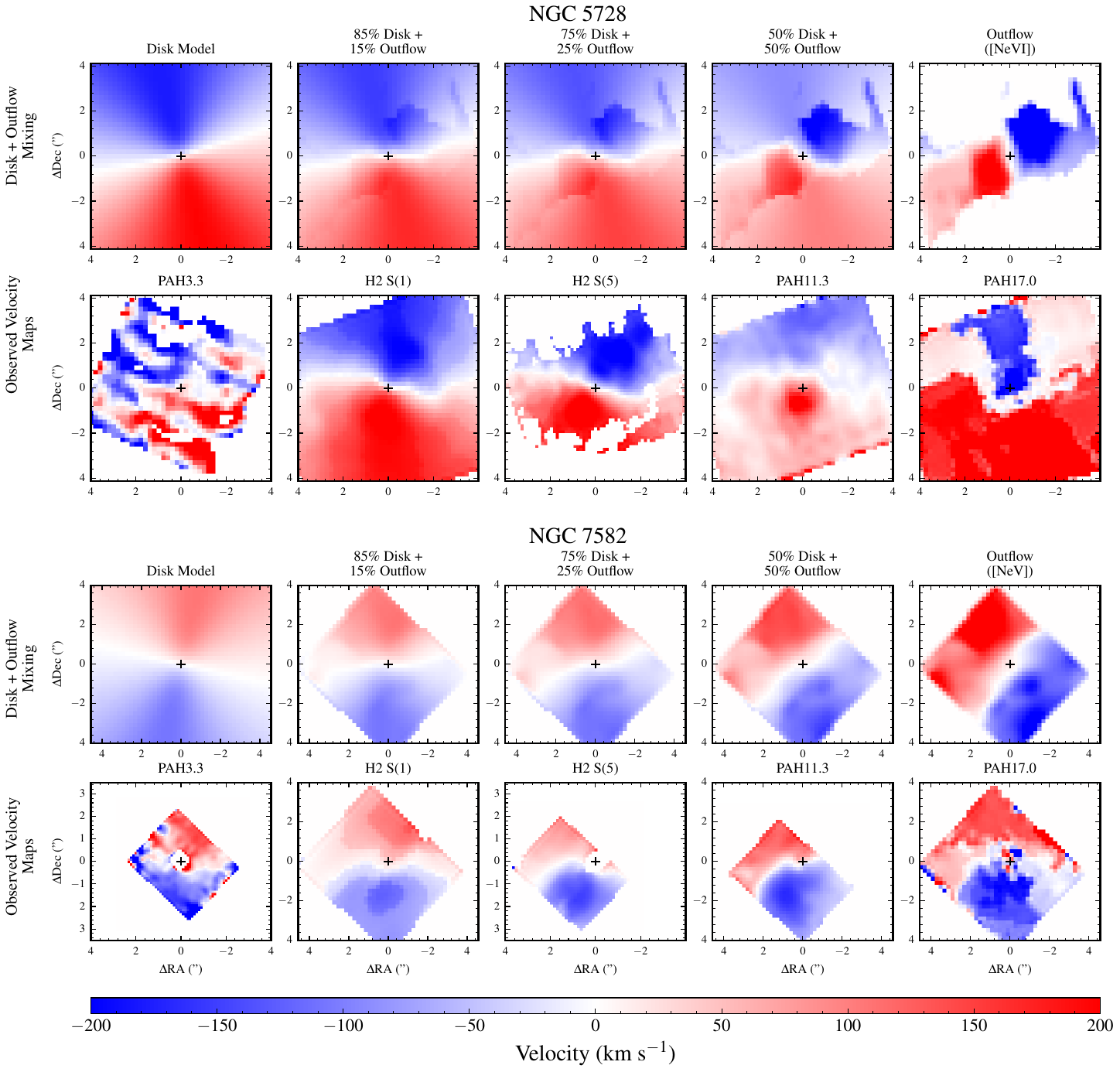}
    \caption{Demonstration of the effect of various contributions of the outflow and disk on the resultant velocity map. The top two panels are for NGC 5728 while the bottom two are for NGC 7582. For each object the top row shows the progression from left to right of the velocity map from pure disk to pure outflow. The lower panels show observed velocity maps for various features where the 3.3 PAH on the left shows pure disk while the other features show some mixture of outflow and disk. }
    \label{fig:OutflowMixing} 
\end{figure*}

We further demonstrate the effect of the contribution of the outflow to the observed velocity maps in Fig. \ref{fig:OutflowMixing}, where we take the pure disk model and add the outflow with progressively higher contributions. We find that the 15$\%$ outflow contribution provides a good match to the H$_2$ S(1) line while the 25$\%$ outflow map matches the hotter $H_2$ S(5) line. for both NGC 5728 and NGC 7582. The 11.3 $\mu$m PAH matches a higher outflow contribution at up to 50 $\%$ for both targets, although for NGC 5728 is missing the blueshifted side as previously discussed. The 17 $\mu$m PAH shows a similar velocity map to the $H_2$ S(5) line, with an outflow contribution of $\sim25\%$, although the map is very noisy for NGC 5728.

\section{Discussion}
\label{sec:Discussion}
\subsection{Dust in AGN Outflows}

From the analysis in section \ref{sec:DiskModelling}, the outflow signature is clear in the velocity map of the 11.3 $\mu$m PAH in both NGC 7582 and NGC 5728. The other PAH features show much weaker evidence of the outflow however the 17.0 $\mu$m PAH also shows evidence of the outflow after subtracting the disk model. We discuss the implications on the properties of the PAH molecules in section \ref{sec:PAHProp}. We do not find any evidence of the outflow in the PAH velocity maps for ESO 420-G13 however. 

The picture is similar for the molecular gas as seen by the rotational transitions of H$_2$. Both NGC 7582 and NGC 5728 show a combination of disk and outflow in the molecular gas, with the higher-J rotational transitions containing a greater contribution from the outflow. The only other target that shows clear evidence for the presence of the outflow in the H$_2$ rotational lines in ESO 137-G034. Unfortunately the signal to noise of the PAH features is low in this target and so we cannot verify if there are PAHs in the outflow. The remaining targets show no clear evidence of either H$_2$ or PAHs in the outflow which does not rule out their presence. The lack of detection may be due to the galaxy being viewed edge-on which is the case for NGC 5506 and NGC 7172 or the outflow kinematics are not distinct from rotation. This is the case for NGC 3227. However ESO 420-G13, NGC 3081 and NGC 4051 show no evidence for the outflow in the kinematics of the rotational H$_2$ lines or the PAHs (where the signal to noise is sufficient).

The correlation between PAH emission and molecular gas content \citep[e.g.][]{Regan2006, Cortzen2019, Shivaei2024b} is well documented, typically attributed to star-forming regions however this is also the case in outflows for both star-formation driven outflows such as in M82, \citep[][]{Beirao2015, Lopez2025, Villanueva2025} but also AGN outflows \citep[e.g.][]{Garcia-Bernete2024b}.

\subsubsection{Acceleration of dust grains}
\begin{figure}
	\includegraphics[width=\columnwidth]{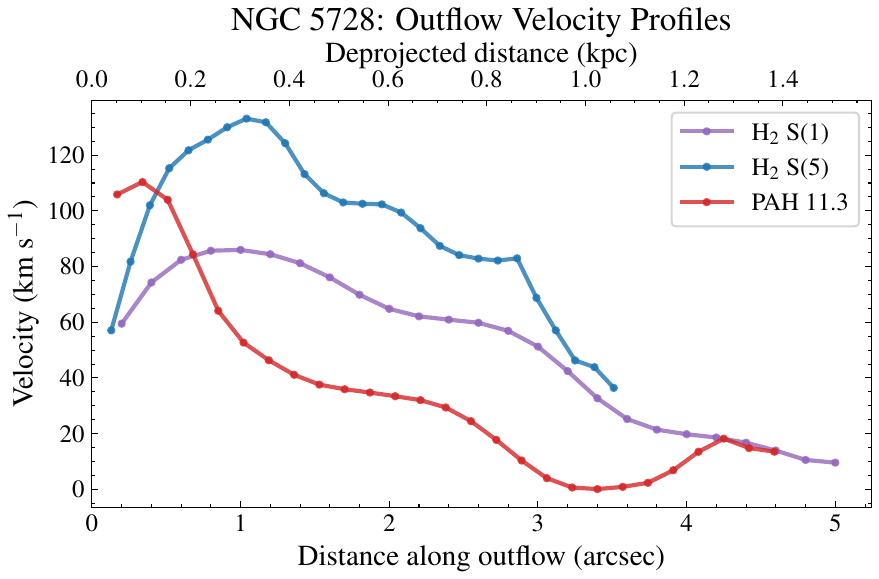}
    \caption{Velocity profiles of the molecular gas via the H$_2$ S(1) and H$_2$ S(5) lines compared to the 11.3 $\mu$m PAH feature. These were extracted after subtracting the disk kinematics and only for the red side of the outflow. }
    \label{fig:VelProf} 
\end{figure}

To investigate the acceleration of the PAHs in the outflow we extract the velocity profile of the 11.3 $\mu$m PAH feature after subtracting the disk kinematics. We plot this in Fig. \ref{fig:VelProf} alongside the molecular gas via the H$_2$ S(5) and S(1) lines. We show the distance as both observe angular distance in arcseconds as well as the deprojected distance assuming an outflow inclination of 49$^{\circ}$ \citep[][]{Shimizu2019}.

The profiles clearly show an initial acceleration before a deceleration out to $\sim1$ kpc. For the molecular gas, the acceleration takes place up to $\sim0.3$ kpc before the deceleration, with the S(5) line reaching tracing higher velocity gas. The PAH feature however shows a quicker acceleration and an earlier deceleration at $\sim0.1$ kpc. This is expected as PAHs have a higher surface area to mass ratio than molecular gas, making them easier to radiatively accelerate close to the AGN \citep[e.g.][]{Fabian2008}, but also easier to decelerate as material is swept up from the ISM into the outflow. This scenario assumes radiative acceleration but considering the relatively larger scales of this outflow ($\gtrsim100$pc), acceleration due to a jet or AGN winds is also plausible. NGC 5728 has a radio detected jet on scales of $>200$pc \citep[][]{Schommer1988, Durre2018} however evidence for a jet in NGC 7582 is more tentative \citep[][]{Forbes1998, Ricci2018}.

The drop in velocity of the PAHs with distance could also be due to destruction due to shocks where as PAHs are accelerated to high speed, they become susceptible to shocks at velocities of $\gtrsim100$ km s$^{-1}$ \citep[e.g.][]{Micelotta2010, Donnan2023b}, causing only slow moving PAHs to remain at higher distances from the AGN. 


\subsubsection{Dust Origin}

With this kinematic evidence for PAHs it is clear that dust forms part of the outflow from AGN however a key question still remains. Does the dust originate from the torus and forms part of the polar dust structure \citep[e.g.][]{Honig2019} or does the dust simply reside in the ISM and is entrained in the ISM of the host galaxy \citep[e.g.][]{Alonso-Herrero2019, Shimizu2019, Garcia-Bernete2021}?

One argument for the latter is from PAH band ratios, where AGN outflows show more neutral PAHs \citep[e.g.][]{Garcia-Bernete2024b} however this only appears to happen when there is strong coupling/interaction between the outflow and the circumnuclear star-forming disk. This is the case for both NGC 5728 and NGC 7582 \citep[e.g.][]{Garcia-Bernete2024b, Garcia-Bernete2026, Veenema2026}. Considering that the outflows we have detected are on larger scales ($\sim$1kpc) than the torus \citep[$\sim10$s pc][]{Garcia-Burillo2019, Garcia-Burillo2021, Combes2019, Alonso-Herrero2021}, suggests that these dusty outflows arise from the interaction of the AGN with its circumnuclear environment.

\begin{figure}
	\includegraphics[width=\columnwidth]{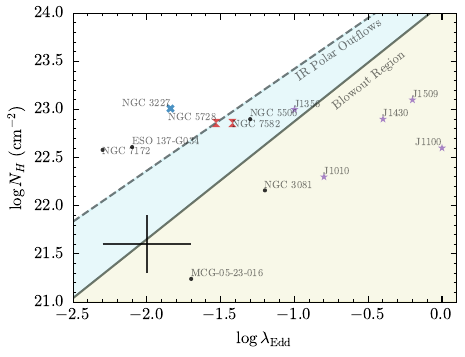}
    \caption{Diagram of the column density, $N_H$, vs the Eddington ratio, $\lambda_{\textrm{Edd}}$. The column densities are inferred from CO in \citet{Garcia-Burillo2024}. The sources that show PAHs in the outflow are marked with a red hourglass marker while the object where we rule out PAHs in the outflow is shown with a blue cross.  Objects where we do not see any evidence for PAHs in the outflows but cannot rule out their presence are shown with the black circles. We also plot the quasars from \citet{RamosAlmeida2025, RamosAlmeida2026} as purple stars for reference. We show the limit from \citet{Fabian2008} (solid line) beyond which the Eddington ratio exceeds the effective Eddington ratio for dust creating the ``Blowout Region''. We also show the limit from \citet{Venanzi2020} (dashed line) where gravity balances the radiation pressure, creating a parameter space where we would expect to see polar outflows. The cross in the bottom left corner shows typical uncertainties in both axes.}
    \label{fig:ColDens} 
\end{figure}

We can further address this question by inspecting the position of the galaxies on the column density, $N_{\rm H}$, vs Eddington ratio, $\lambda_{\rm Edd}$ diagram. This was first introduced by \citet{Fabian2008}, where the effective Eddington limit for dust is lower than the gas, resulting in a region of the $N_{\rm H}$ vs $\lambda_{\rm Edd}$ space where the AGN clear all of their dust \citep[e.g.][]{Ricci2017b}. For the GATOS sample, this kind of diagram was first shown in \citet{Alonso-Herrero2021}.
We determine the hydrogen column density from CO observations from \citet{Garcia-Burillo2024}, assuming a Milky Way $\alpha_{\textrm{CO}}$ and CO line ratios consistent with the highest spatial resolution data of NGC 1068 \citep[][]{Garcia-Burillo2014, Viti2014}. We plot the hydrogen column density against the Eddington ratio in Fig. \ref{fig:ColDens}. It is worth noting that there is a lot of uncertainty in measuring the column density as different CO lines and/or X-ray data will give different results. We therefore show a wide error range in Fig. \ref{fig:ColDens}. Moreover, the line of sight can affect the measured value of $N_{\rm H}$. We show the theoretical regions for polar dust outflows from \citet{Venanzi2020} and the blowout region from \citet{Fabian2008}.

We find that the two sources that show PAHs in the outflow, NGC 5728 and NGC 7582, are consistent with the polar outflow region of the plot, although the large errors should be noted. The other sources are spread throughout the parameter space with NGC 3227, which we rule out showing PAHs in the outflow, appearing outwith the polar outflow region. This may suggest that the outflows originated from the torus however the presence of coupling in the two sources that show PAHs in the outflow also suggests that the interaction of the outflow with the circumnuclear ISM may also be necessary to entrain material into the outflow.

A more observationally driven parameter space, presented in Figure 2 of \citet{Garcia-Burillo2024}, is the molecular gas concentration vs X-ray luminosity, where the gas concentration is the ratio of the molecular surface density at 50 pc to 200 pc. In this plot, two regions are observed, the ``AGN build-up'' where at low X-ray luminosity the gas concentration increases and the ``AGN feedback'' branch, where at high X-ray luminosity the gas concentration decreases. This diagram probes larger scales than the $N_{\rm H}$, vs $\lambda_{\rm Edd}$, up to 200 pc, where all the sources in our sample are in the ``Feedback Branch''. It is therefore unsurprising to find objects that show dusty outflows in this region of parameter space.

\subsection{PAH Properties}
\label{sec:PAHProp}

\begin{figure}
	\includegraphics[width=\columnwidth]{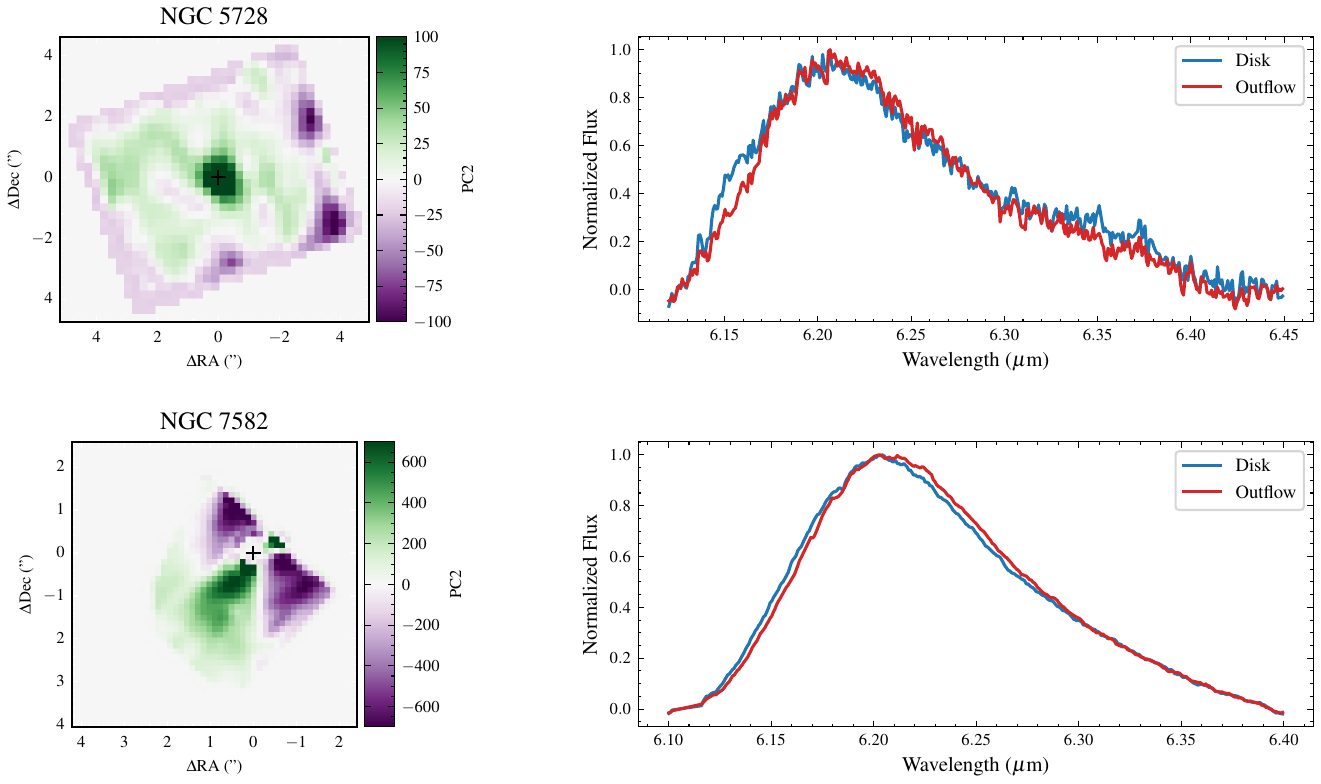}
    \caption{The second principal components of the 6.2 $\mu$m PAH features for NGC 5728 and NGC 7582. The left panels show the tomograms while the right panels show the PAH profiles for regions where the tomogram is either positive or negative, corresponding to different intrinsic profiles in the galaxy disk or outflow. The right panels are not the eigenspectra but rather the average profile in the outflow region vs the disk as defined by the tomograms in the left panel, where either positive or negative values indicate disk or outflow.
    We use a different colourscale to differential from velocity.}
    \label{fig:6.2PCA} 
\end{figure}

The properties of PAHs are known to be altered in the vicinity of AGN outflows \citep[][]{ Garcia-Bernete2022b, Garcia-Bernete2022c, Garcia-Bernete2024b, Rigopoulou2024, Zhang2024}, in particular appearing more neutral, signified by a relatively weaker 6.2 $\mu$m PAH emission. In this work, we found the 6.2 $\mu$m feature challenging to obtain any kinematics as the signal to noise is relatively low in the outflow region, potentially due to preferential destruction of ionised grains \citep[e.g.][]{Garcia-Bernete2024b}, making extracting kinematics difficult. Secondly, the PCA technique is sensitive to differences in the intrinsic PAH profile, and cannot necessarily differentiate between kinematics affecting the observed profile or a different intrinsic profile. In the case of the 6.2 $\mu$m PAH, considering this feature traces PAHs heavily altered by the outflow, it makes sense that the intrinsic profile may be different in the outflow region. We show the different intrinsic profiles, albeit only with subtle differences, as detected in the second principal component for NGC 5728 and NGC 7582 in Fig. \ref{fig:6.2PCA}. These intrinsic profiles, as shown in the right panels of Fig. \ref{fig:6.2PCA}, are calculated as the average PAH profile over spaxels where the tomogram of the second principal component, PC2, is either positive or negative. For example, we observe that the value of PC2 is negative in outflow regions in NGC 7582 but is positive in the disk. Therefore the ``outflow'' profile is calculated as the average PAH profile across spaxels where the tomogram is negative.

We confirm that the PCA analysis of the 6.2 $\mu$m feature is detecting different intrinsic profiles rather than Doppler shifts from kinematics in Appendix \ref{sec:IntProf}, where we can inspect the shape of the eigenspectra to see if it is consistent with a Doppler shift. We find that the 11.3 $\mu$m PAH is robust at tracing kinematics while the 6.2 $\mu$m feature is not.

Interestingly, as presented in section \ref{sec:DiskModelling} and
as shown in Fig. \ref{fig:DiskModellingNGC7582} and Fig. \ref{fig:OutflowMixing}, the 3.3 $\mu$m PAH purely traces rotation in the disk and shows no contribution from the outflow compared to the 11.3 $\mu$m PAH and 17 $\mu$m PAH. This suggests that small PAH molecules are also destroyed in the outflows of AGN. Small PAHs are thought to be easier to destroy than large PAHs \citep[e.g.][]{Lebouteiller2007, O'Dowd2009, Micelotta2010, Zhen2015,  Lange2025}, suggesting that the radiation field of the AGN preferentially destroys small grains.

Large and neutral PAH molecules are consistent with PAH band ratios measured for Seyfert galaxies \citep[][]{Garcia-Bernete2022b, Garcia-Bernete2022c, Garcia-Bernete2024b, Donnelly2024}, where the ionised and small PAH molecules are thought to be destroyed by the radiation field of the AGN. The mechanism by which this takes place however is not fully understood but it may be related to ionised PAHs having weaker carbon skeletons due to higher Coulomb forces when electrons are lost or gained through ionisation \citep[][]{Leach1986, Voit1992}, making them more susceptible to photo-destruction. 

The lack of kinematics of small and ionised PAHs in the outflow provides new, corroborating evidence in addition to the band ratios, that PAHs are altered by AGN outflows becoming more neutral and larger.

\subsection{Aliphatics in the Outflow?}
The 3.4 $\mu$m feature results from aliphatic hydrocarbons, where instead of a closed carbon chain (aromatic), open carbon chains, either as side groups or as independent molecules, produce the feature at 3.4 $\mu$m \citep[][]{Wexler1967, Duley1981, Pauzat1999}.

\begin{figure}
	\includegraphics[width=\columnwidth]{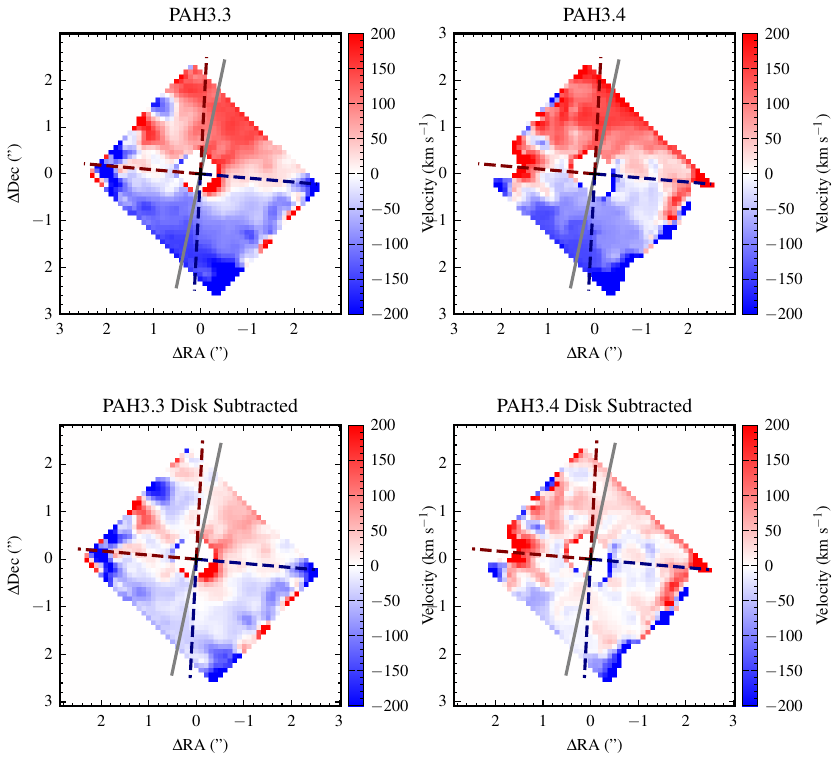}
    \caption{Top: The velocity maps of the 3.3 $\mu$m PAH feature and the 3.4 $\mu$m aliphatic feature for NGC 7582. Bottom: Disk subtracted velocity maps of the 3.3 $\mu$m PAH feature and the 3.4 $\mu$m aliphatic feature. The major axis of the disk is shown with the solid grey line while the outflow cone is shown with the dashed lines in red and blue for the redshift and blueshift sides of the cone respectively.}
    \label{fig:PAH34} 
\end{figure}

Given that the 3.4 $\mu$m feature is relatively weak compared to the 3.3 $\mu$m PAH feature, the signal to noise of most targets in this work is not sufficient to produce a velocity map. However for NGC 7582 we are able to produce a velocity map which is shown in Fig. \ref{fig:PAH34} in comparison with the 3.3 $\mu$m PAH. We also detect the double peak in the 3.4 $\mu$m feature at 3.395 $\mu$m and 3.405 $\mu$m. The strong presence of the latter feature indicates that the PAHs are shielded by molecular gas \citep[][]{Thatte2026} as it arises from more fragile aliphatic chains which are more easily destroyed in hard radiation fields.


We find no convincing evidence for the presence of the outflow in the residual velocity map of the 3.4 $\mu$m feature in Fig. \ref{fig:PAH34} with the velocity well reproduced by the disk model. The 3.4 $\mu$m feature is thought to be dominated by aliphatic side groups on aromatic molecules \citep[][]{Joblin1996, Pendleton2002, Yang2023} and so it is not necessarily a tracer of the total aliphatic content. Therefore the lack of 3.4 $\mu$m in the outflow could be explained as the stripping of aliphatic side groups from large PAHs in the hard radiation field of the outflow \citep[][]{Joblin1996, Jones2013, Yang2016}.


\section{Conclusions}
In this work we presented the kinematics of PAH features using PCA tomography for 10 nearby Seyfert galaxies as part of the GATOS sample. We present the first kinematic detection of outflowing dust from AGN. Our main findings are
\begin{itemize}
    \item We find that the velocity field of the 3.3 $\mu$m PAH is purely due to rotation in the circumnuclear disk of the host galaxy for all our targets where 3.3 $\mu$m PAH is detected.  
    \item We find that the 11.3 $\mu$m PAH and 17.0 $\mu$m PAH show contribution from the outflow in 2 out of the 10 targets, namely NGC 5728 and NGC 7582. After subtracting a disk model, the residuals show clear evidence of the outflow.
    \item The 6.2 $\mu$m PAH fails to produce a velocity map from the PCA tomography technique, with instead the PCA showing an alternative intrinsic profile in the outflow regions suggesting the ionised PAHs are altered in AGN outflows. This fact as well as the lack of 3.3 $\mu$m PAH in the outflow is consistent with band ratios that suggest AGN outflows contain more neutral and larger PAHs than star-forming regions. 
    \item We find that the objects that show PAHs in the outflow also show outflow contributions in the molecular gas as traced by the H$_2$ rotational transitions. This suggests that the PAHs are correlated with the molecular gas in AGN outflows.
    \item We find a lack of evidence for the 3.4$\mu$m aliphatic feature in AGN outflows, consistent with the idea of stripping of aliphatic side groups in the hard radiation fields of AGN outflows.
\end{itemize}
We suggest that the presence of PAHs in AGN outflows is the result of the interaction and coupling to the circumnuclear environment where material is entrained into the outflow.
This kind of work presents the first kinematic analysis of a new phase of AGN outflows. Expanding the sample to cover a wider parameter space will enable us to understand how the dusty phase of AGN outflows which is potentially linked to how AGN evolve and clear their obscuring dust.

\section*{Acknowledgements}
FRD acknowledges support from STFC through studentship ST/W507726/1. IGB is supported by the Programa Atracci\'on de Talento Investigador ``C\'esar Nombela'' via grant 2023-T1/TEC-29030 funded by the Community of Madrid, and acknowledges support from the research project PID2024-159902NA-I00 funded by the Spanish Ministry of Science and Innovation/State Agency of Research (MCIN/AEI/10.13039/501100011033) and FSE+. DR acknowledge support from STFC through grant ST/S000488/1 and ST/W000903/1. RAR acknowledges the support from the Conselho Nacional de Desenvolvimento Científico e Tecnológico (CNPq; Projects 303450/2022-3, and 403398/2023-1), the Coordenação de Aperfeiçoamento de Pessoal de Nível Superior (CAPES; Project 88887.894973/2023-00), and Fundação de Amparo à Pesquisa do Estado do Rio Grande do Sul (FAPERGS; Project 25/2551-0002765-9). JAFO acknowledges financial support by the Spanish Ministry of Science and Innovation (MCIN/AEI/10.13039/501100011033), by ``ERDF A way of making Europe'' and by ``European Union NextGenerationEU/PRTR'' through the grants PID2021-124918NB-C44 and CNS2023-145339; MCIN and the European Union -- NextGenerationEU through the Recovery and Resilience Facility project ICTS-MRR-2021-03-CEFCA. MS acknowledges support by the Ministry of Science, Technological Development and Innovation of the Republic of Serbia (MSTDIRS) through contract no. 451-03-136/2025-03/200002 with the Astronomical Observatory (Belgrade). CRA and AA thank the Agencia Estatal de Investigacion of the Ministerio de Ciencia, Innovacion y Uni270 versidades (MCIU/AEI) under the grant “Tracking active galactic nuclei feedback from parsec to kiloparsec scales”, with reference PID2022-141105NB-I00 and the European Regional Development Fund (ERDF). AA also acknowledges support from the European Union (WIDERA ExGal-Twin, GA 101158446). CR acknowledges support from SNSF Consolidator grant F01$-$13252, Fondecyt Regular grant 1230345, ANID BASAL project FB210003 and the China-Chile joint research fund. SGB acknowledges support from the Spanish grant PID2022-138560NB-I00, funded by MCIN/AEI/10.13039/501100011033/FEDER, EU. MPS acknowledges support from grants RYC2021-033094-I, CNS2023-145506, and PID2023-146667NB-I00 funded by MCIN/AEI/10.13039/501100011033 and the European Union NextGenerationEU/PRTR. AJB acknowledges funding from the “FirstGalaxies” Advanced Grant from the European Research Council (ERC) under the European Union’s Horizon 2020 research and innovation program (Grant agreement No. 789056). EB acknowledges support from the Spanish grants PID2022-138621NB-I00 and PID2021-123417OB-I00, funded by MCIN/AEI/10.13039/501100011033/FEDER, EU. AAH and LHM acknowledge support from grant PID2021-124665NB-I00  funded by MCIN/AEI/10.13039/501100011033 and by ERDF A way of making Europe. O.G.-M. acknowledges the support received by the UNAM DGAPA-PAPIIT project IN109123 and SEHCITI Ciencia de Frontera project CF-2023-G100. SFH acknowledges support through UK Research and Innovation (UKRI) under the UK government’s Horizon Europe Funding Guarantee (EP/Z533920/1, selected in the 2023 ERC Advanced Grant round) and an STFC Small Award (ST/Y001656/1)

\section*{Data Availability}
This work is based [in part] on observations made with the NASA/ESA/CSA James Webb Space Telescope. The data were obtained from the Mikulski Archive for Space Telescopes at the Space Telescope Science Institute, which is operated by the Association of Universities for Research in Astronomy, Inc., under NASA contract NAS 5-03127 for JWST. These observations are associated with programs GO 5017, GO 1875, GO 1670, GO 3335. The data are downloadable from the \href{https://mast.stsci.edu/portal/Mashup/Clients/Mast/Portal.html}{MAST archive}.



\bibliographystyle{mnras}
\bibliography{References} 

\section*{}
\rule{\linewidth}{0.4pt}  
$^{1}$Department of Physics, University of Oxford, Keble Road, Oxford, OX1 3RH, UK\\
$^{2}$Department of Astrophysics, University of California San Diego, 9500 Gilman Drive, San Diego, CA 92093, USA\\
$^{3}$Centro de Astrobiolog\'{\i}a (CAB), CSIC--INTA, Camino Bajo del Castillo s/n, E--28692 Villanueva de la Ca\~nada, Madrid, Spain\\
$^{4}$School of Sciences, European University Cyprus, Diogenes Street, Engomi, 1516 Nicosia, Cyprus\\
$^{5}$Instituto de Astrof\'isica de Canarias, Calle Vía Láctea, s/n, E-38205, La Laguna, Tenerife, Spain\\
$^{6}$Departamento de Astrof\'isica, Universidad de La Laguna, E-38206, La Laguna, Tenerife, Spain\\
$^{7}$Departamento de F\'isica de la Tierra y Astrof\'isica, Fac. de CC. F\'isicas, Universidad Complutense de Madrid, 28040 Madrid, Spain\\
$^{8}$Instituto de F\'isica de Part\'iculas y del Cosmos IPARCOS, Fac. CC. F\'isicas, Universidad Complutense de Madrid, 28040 Madrid, Spain\\
$^{9}$School of Mathematics, Statistics and Physics, Newcastle University, Newcastle upon Tyne NE1 7RU, UK\\
$^{10}$LUX, Observatoire de Paris, Coll\`ege de France, PSL University, CNRS, Sorbonne University, Paris, France\\
$^{11}$Max Planck Institute for Extraterrestrial Physics (MPE), Giessenbachstr. 1, 85748 Garching, Germany\\
$^{12}$Institute of Astrophysics, Foundation for Research and Technology-Hellas (FORTH), Heraklion, GR-70013, Greece\\
$^{13}$Centro de Estudios de F\'isica del Cosmos de Arag\'on (CEFCA), Plaza San Juan 1, E--44001 Teruel, Spain\\
$^{14}$Department of Physics \& Astronomy, University of Southampton, Highfield, Southampton, SO17 1BJ, UK\\
$^{15}$Observatorio Astron\'omico Nacional (OAN--IGN), Observatorio de Madrid, Alfonso XII, 3, 28014 Madrid, Spain\\
$^{16}$Instituto de Radioastronomía y Astrofísica (IRyA), Universidad Nacional Autónoma de México, Morelia, Michoacán, Mexico\\
$^{17}$Department of Physics \& Astronomy, University of Alaska Anchorage, Anchorage, AK 99508-4664, USA\\
$^{18}$National Astronomical Observatory of Japan, National Institutes of Natural Sciences (NINS), 2-21-1 Osawa, Mitaka, Tokyo 181-8588, Japan\\
$^{19}$Department of Astronomy, School of Science, Graduate University for Advanced Studies (SOKENDAI), Mitaka, Tokyo 181-8588, Japan\\
$^{20}$Telespazio UK for ESA, ESAC, Camino Bajo del Castillo s/n, 28692 Villanueva de la Cañada, Spain\\
$^{21}$Space Telescope Science Institute, 3700 San Martin Drive, Baltimore, MD 21218, USA\\
$^{22}$Instituto de F\'isica Fundamental, CSIC, Calle Serrano 123, 28006 Madrid, Spain\\
$^{23}$Department of Astronomy, University of Geneva, ch. d’Ecogia 16, 1290, Versoix, Switzerland\\
$^{24}$Instituto de Estudios Astrofísicos, Facultad de Ingeniería y Ciencias, Universidad Diego Portales, Av. Ejército Libertador 441, Santiago, Chile\\
$^{25}$Departamento de F\'isica, CCNE, Universidade Federal de Santa Maria, Av. Roraima 1000, 97105--900 Santa Maria, RS, Brazil\\
$^{26}$LESIA, Observatoire de Paris, PSL Research University, CNRS, Sorbonne Université, Paris-Cité University, Meudon, France\\
$^{27}$Astronomical Observatory, Volgina 7, 11060 Belgrade, Serbia\\
$^{28}$Sterrenkundig Observatorium, Universiteit Gent, Krijgslaan 281-S9, B-9000 Gent, Belgium\\
$^{29}$Department of Physics and Astronomy, The University of Texas at San Antonio (UTSA), 1 UTSA Circle, San Antonio, TX 78249-0600, USA\\





\appendix

\section{Spectra}
\begin{figure*}
	\includegraphics[width=\textwidth]{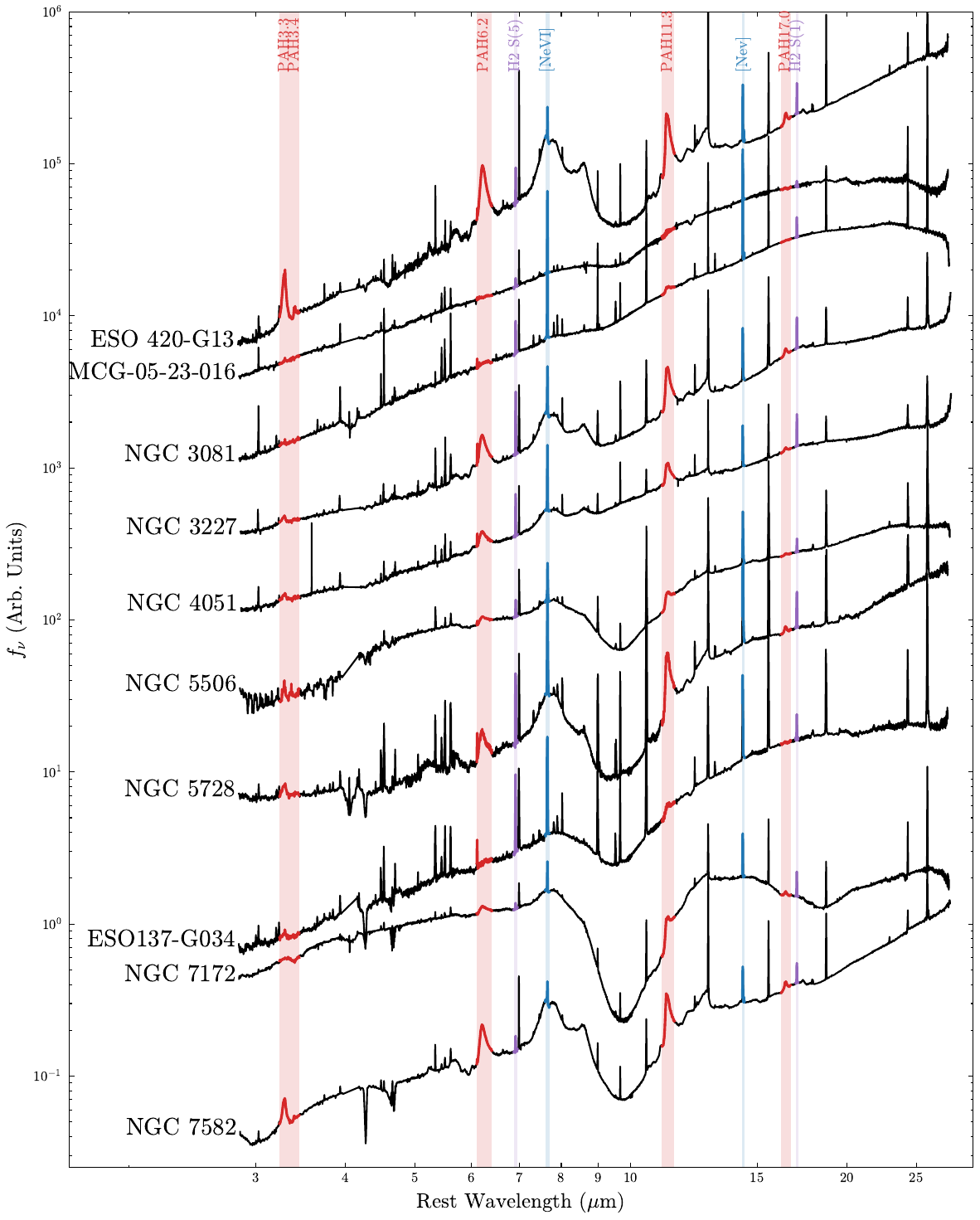}
    \caption{Integrated spectra of all 10 Seyfert galaxies used in this work, extracted from the NIRSpec and MIRI cubes through a wide, 2'' radius aperture. We highlight the emission features used to create velocity maps with the PAHs in red, ionised gas in blue and molecular gas in purple. The spectra have been arbitrarily multiplicatively scaled in flux to make it easy to visualise.}
    \label{fig:Spectra} 
\end{figure*}

We show integrated spectra of all the Seyfert galaxies analysed in this work in Fig. \ref{fig:Spectra}. We extracted the spectra through a circular aperture with a radius of 2'' and thus the spectra contain contribution from both the AGN and the circumnuclear environment. 

\section{Comparison with Line Fitting}
\label{sec:LineTest}
To ensure that the velocity maps produced by the PCA technique is robust we compare it against fitting a Gaussian to each spaxel for the emission lines where such analysis can be done. For each spaxel we fit a Gaussian where the amplitude, width and central wavelength are allowed to vary. The velocity for each spaxel is determined from the fitted central wavelength compared to the rest frame wavelength. 

\begin{figure}
	\includegraphics[width=\columnwidth]{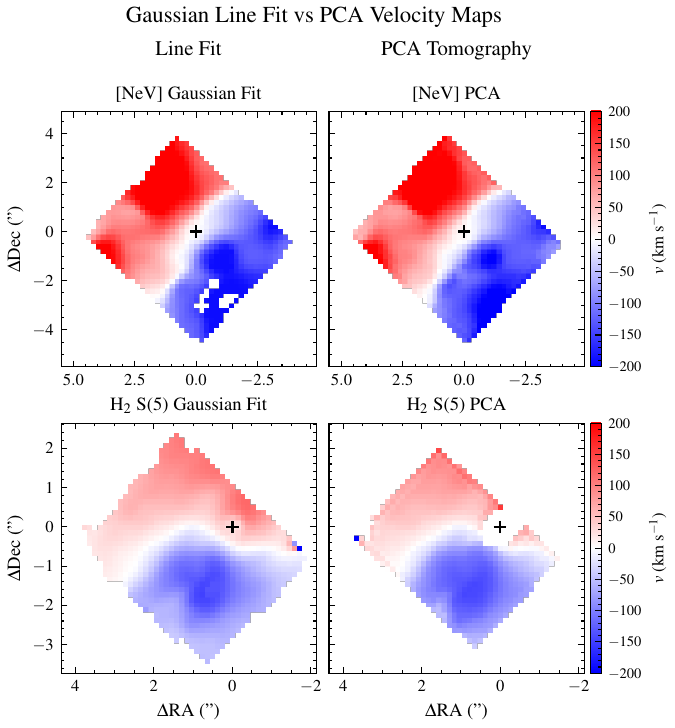}
    \caption{Comparison between a more traditional line fitting to produce velocity maps and the PCA technique used in this work for NGC 7582, where we show the [NeV] and H$_2$ S(5) line. The test shows that the PCA technique is robust and able to produce accurate velocity maps. }
    \label{fig:LineTest} 
\end{figure}

We show this test for NGC 7582 for [NeV] tracing the outflow and H$_2$ S(5) which contains a mixture of disk and outflow in Fig. \ref{fig:LineTest} where the line fits clearly match the velocity maps inferred using the PCA technique.

\section{Intrinsic profile variations or kinematics }
\label{sec:IntProf}
We perform an additional check on whether the PCA decomposition is detecting variations in the emission profile due to Doppler shifts from motion or a different intrinsic profile due to the properties of the PAHs and/or differences in excitation. We can do this by inspecting the shape of the eigenspectrum of the second principal component.

A Doppler shift of velocity, $v$, of a PAH feature, $F_{\rm rest}\left(\lambda'\right)$, can be written as 
\begin{equation}
    f_{\rm obs}\left(\lambda\right) = f_{\rm rest}\left[\lambda\left(1+\frac{v}{c}\right)\right],
\end{equation}
where $c$ is the speed of light and $\lambda' = \lambda\left(1+\frac{v}{c}\right)$. We can rewrite this expression using a Taylor expansion as $v\ll c$ giving 
\begin{equation}
    f_{\rm obs}\left(\lambda\right) \approx f_{\rm rest}\left(\lambda\right) + \frac{v}{c}\frac{df_{\rm rest}}{d\lambda} \lambda .
\end{equation}
allowing one to write a Doppler shift as a linear function. The first term is the rest frame emission of the feature while the second term is proportional to the 1\textsuperscript{st} derivative of the first term. Considering that the PCA decomposition is linear, we can see that the first principal component should contain the rest frame emission while the second principal component should be proportional to the derivative of the first in the case of a Doppler shift. If instead the intrinsic profile varies, we would expect extra power in the one/both of the wings or the shoulder of the feature rather than a clean shift in wavelength that a Doppler shift produces. Therefore, this gives us a powerful discriminant on whether the changes in the observed emission profile of the PAHs are truly due to velocity shifts or instead are due to variations in the intrinsic profile by inspecting the shape of the eigenspectrum of the second principal component.

\begin{figure}
	\includegraphics[width=\columnwidth]{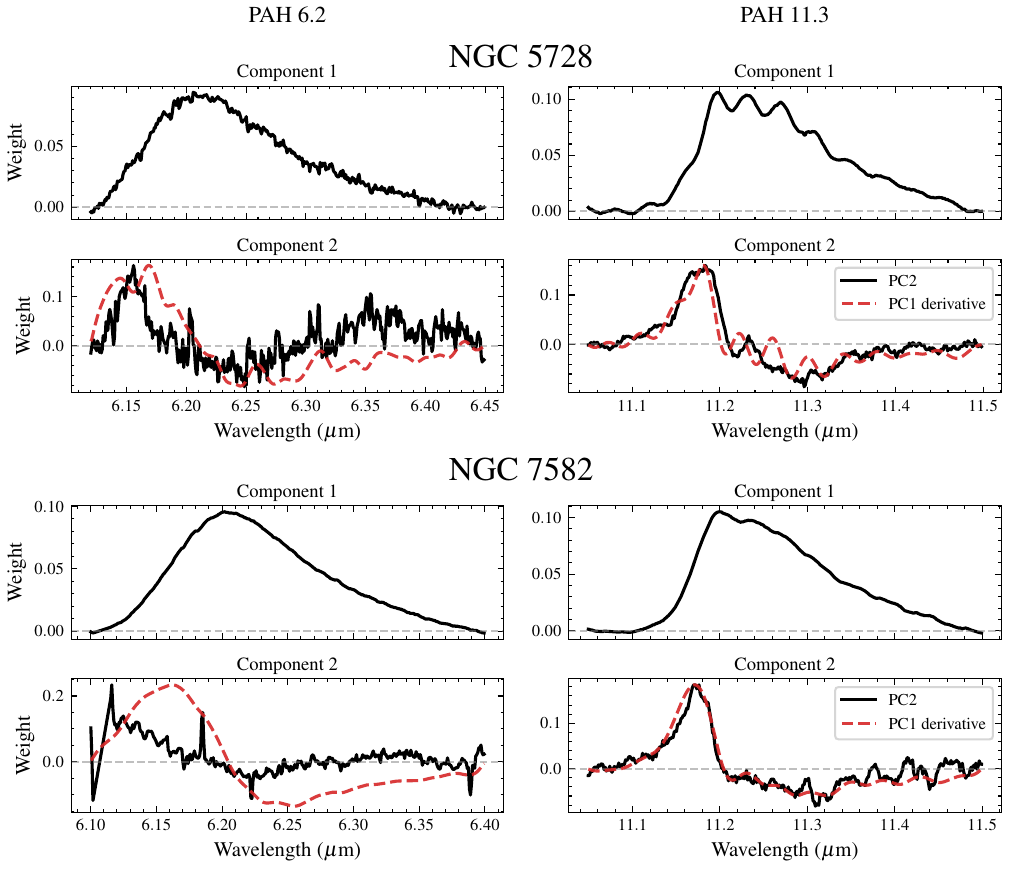}
    \caption{The eigenspectra of the first and second principal components of the 6.2 $\mu$m and 11.3 $\mu$m for NGC 5728 and NGC 7582. The red dashed line shows the derivative of the eigenspectra of the first principal component which should match the eigenspectra of the second component in the case of a Doppler shift. This is indeed the case for the 11.3 $\mu$m feature, making it a robust tracer of PAH kinematics in these objects but not the 6.2 $\mu$m feature. Note that there is some fringing still present in the data, particularly visible in component 1 of the 11.3$\mu$m feature for NGC 5728.}
    \label{fig:Eigen} 
\end{figure}

As presented in this work, the 11.3 $\mu$m PAH appears to trace velocity only whereas the 6.2 $\mu$m PAH shows variations in the intrinsic profile as demonstrated in Fig. \ref{fig:6.2PCA}, preventing one from inferring the velocity from the 6.2 $\mu$m feature. We can see this clearly by inspecting the eigenspectrum of the second principal component for both features for the two galaxies where we find kinematic evidence for the 11.3 $\mu$m PAH feature in their respective AGN outflows as shown in Fig. \ref{fig:Eigen}. We find that for both galaxies, the eigenspectrum of the second principal component of the 11.3 $\mu$m PAH feature is consistent with the derivative of the first meaning we are indeed tracking kinematics of the PAHs with the 11.3 $\mu$m PAH feature. 

The eigenspectrum of the second principal component of the 6.2 $\mu$m feature is not consistent with the derivative of the first for either galaxies which suggests there is sone difference in the intrinsic profile of the 6.2 $\mu$m PAH feature rather than kinematics. This test reinforces what we see in the tomograms as discussed in section \ref{sec:PAHProp}, where it is clear that there is altering of the 6.2 $\mu$m PAH profile in the outflow regions of NGC 5728 and NGC 7582 compared to the disk.

\bsp	
\label{lastpage}
\end{document}